%% file: preprint_sissa.tex
\documentclass{aa}
\usepackage{graphicx}
\newcommand{\Msun}{\mbox{$M_{\odot}$}}
\newcommand{\Msolar}{\mbox{$M_{\odot}$}}
\newcommand{\sub}[1]{\mbox{$_{\rm #1}$}}

\newcommand{\Mto}{\mbox{$M\sub{TO}$}}

\newcommand{\Mcore}{\mbox{$M\sub{c}$}}

\newcommand{\diff}{\mbox{d}}
\newcommand{\beq}{\begin{equation}}
\newcommand{\eeq}{\end{equation}}
\newcommand{\beqa}{\begin{eqnarray}}
\newcommand{\eeqa}{\end{eqnarray}}
\newcommand{\benu}{\begin{enumerate}}
\newcommand{\eenu}{\end{enumerate}}
\newcommand{\bite}{\begin{itemize}}
\newcommand{\eite}{\end{itemize}}
\newcommand{\bdes}{\begin{description}}
\newcommand{\edes}{\end{description}}
\newcommand{\comment}[1]{}

\begin{document}

        \title{Coupling emitted light and chemical yields from stars:\\
	a basic constraint to population synthesis models of galaxies}

	\author{Paola Marigo\inst{1,2} \and L\'eo Girardi\inst{2,1}}
	\institute{
Max-Planck-Institut f\"ur Astrophysik, Karl-Schwarzschild-Str.\
        1, D-85741 Garching bei M\"unchen, Deutschland 
	\and
Dipartimento di Astronomia, Universit\`a di Padova,
        Vicolo dell'Osservatorio 2, I-35122 Padova, Italia }


	\offprints{Paola Marigo \\ 
e-mail: marigo@pd.astro.it }

	\date{Accepted for publication in A\&A}

	\abstract{
In this paper we emphasize the close connection between the chemical
and spectrophotometric evolution of stellar systems: 
Chemical yields from stars correspond to a precise fraction of their
emitted light.
We translate this concept quantitatively. 
Starting from simple stellar populations, 
we derive useful analytical relations to calculate the stellar
fuel consumption (emitted light) as a function of basic quantities
predicted by stellar models, i.e. the mass of the core and the
chemical composition of the envelope.
The final formulas explicate the relation between integrated
light contribution (total or limited to particular evolutionary
phases), chemical yields and stellar remnants. We
test their accuracy in the case of low- and intermediate-mass stars,
and indicate the way to extend the analysis to massive stars.
This formalism provides an easy tool to check the internal consistency
between the different stellar inputs adopted in galaxy models: The
fuel computed by means of the analytical formulas (corresponding to a
given set of chemical yields) should be compared to the exact values
given by the luminosity integration along the stellar evolutionary
tracks or isochrones (corresponding to a given set of
spectrophotometric models).  Only if both estimates of the fuel are
similar, the stellar inputs can be considered self-consistent in
terms of their energetics.
This sets an important requirement to galaxy models, also in
consideration of the fact that different sources of input stellar data
are frequently used to model their spectro-photometric and chemical
evolution.
\keywords{stars: evolution -- stars: abundances -- galaxies: evolution
-- galaxies: abundances}
	}  

        \titlerunning{Coupling emitted light and chemical yields from stars}
        \authorrunning{Marigo \& Girardi}

	\maketitle

\section{Introduction}
\label{sec_intro}

Galaxy evolution models represent fundamental tools for the
interpretation of a wide variety of observational data, going from
star counts in our Galaxy to galaxy counts in deep fields.  One of the
essential assumptions in these models is that {\em we know how stars
convert their nuclear fuel into light and newly-synthesized elements}.
Then, this knowledge is used to model the light emission from
galaxies, as well as the chemical enrichment of the galaxy medium and
the baryonic mass locked into stellar remnants. The basic formalism is
fully described in the pioneering work by Tinsley (1980).

However, it is also a fact that light and chemical evolution of
galaxies have been historically developed into two separate
disciplines, that we refer to by the generic names of
``spectro-photometric models'' (SPM) and ``chemical evolution models''
(CEM) of galaxies. This separation has probably been the result of the
very different basic tools and input data required in these two
fields, namely: stellar theoretical isochrones and spectra for SPMs,
and stellar chemical yields for CEMs.  Other distinctive features of
these fields are that in SPMs star formation rates and age-metallicity
relations are usually inputs to the models, whereas in CEMs they are
outputs of the calculations.

This picture is nowadays rapidly changing, since the present trend is
to model complex stellar aggregates like galaxies accounting
simultaneously for all their aspects, mainly spectrophotometric and
chemical, and possibly including the dynamical one. Some examples of
unified SPMs and CEMs of galaxies can be found in Arimoto \& Yoshii
(1986, 1987), Bressan et al.\ (1994), Einsel et al.\ (1995), Vazdekis
et al.\ (1996), Chiosi et al.\ (1998), and Ferreras \& Silk (2000).
The simplest way to couple CEM and SPM is quite obvious: one may
simply re-direct the output from CEM as input to SPM and, as a matter
of fact, this choice has been frequently made in literature. In this
way the models naturally succeed in reproducing some observed scaling
relations, like for instance the colour--magnitude (mass--metallicity)
relation of ellipticals (e.g. Bressan et al 1994).

In other recent astrophysical problems, coupling CEM and SPM may
become a must. For instance, the recent indications by Ibata et al.\
(2000; see also Chabrier 1999) that a significant fraction of the
detected massive compact halo objects (MACHO) could consist of white
dwarfs, would pose non-trivial problems to present
chemo-spectrophotometric models of galaxies: high rates of white dwarf
formation at early epochs would not be compatible with the observed
deep galaxy counts (Charlot \& Silk 1996), and with the modest
chemical enrichment now observed (Fields et al.\ 2000).  Clearly,
these studies point at a link between the light of distant galaxies,
and chemical abundances and stellar remnants in the local universe. In
this regard, we should also mention the development of more ``global''
analyses (e.g. Madau \& Pozzetti 2000; Fall 2001), that consider the
constraints provided by the cumulative emission of cosmic structures
(the optical to far-infrared extragalactic background light), and the
average metallicity and stellar mass density of the present universe.

However, the fact that nowadays different aspects of galaxy evolution
are being considered simultaneously, does not necessarily imply that
the resulting models are self-consistent as a whole. A potential
source of inconsistency could reside in the (quite frequent) use of
different sources of stellar data as input to SPM (e.g. luminosities
and lifetimes) and CEM (e.g.  chemical yields and remnant masses).

In this context, the present paper emphasizes the close relation
between the chemical and spectrophotometric evolution of galaxies.
We derive the expected relations between the integrated light of
a stellar population, the chemical yields of their component stars,
and the mass of stellar remnants.  These provide {\em important
constraints to any model of galaxy evolution}, which have so far been
either not explicitly checked for, or simply neglected.

We propose such a consistency check may be carried out with the aid of
an analytical formalism, based on simple relations involving stellar
parameters.  The check can be performed as follows.  First one derives
the fuel (emitted light) directly from the luminosities of the adopted
stellar tracks or isochrones (Sect.~\ref{sec_ssp}), and then compares
it with the fuel analytically derived (Sect.~\ref{sec_fuel}), which
corresponds to the particular set of chemical yields in use.  Of
course, the test is successful if the two determinations converge to
the same result, as fully discussed in Sect.~\ref{sec_examp}.

\section{ Single-burst stellar populations}
\label{sec_ssp}

Single-burst stellar populations (SSPs) are the building blocks of
the composite populations we call galaxies. The integrated stellar light
in the pass-band $\lambda$ of an SSP of age $t$ is given by
\footnote{Effects such as extinction, gas emission, etc. can be considered
a posteriori in the models.}
        \beq
L_{\lambda}^{\rm SSP}(t) = \int_{0}^{\infty} \phi(M_{\rm i})\,
        L_{\lambda t}(M_{\rm i})\, \diff M_{\rm i}
        \label{eq_isochrone}
        \eeq
where $M_{\rm i}$ corresponds to the initial stellar mass, $L_{\lambda
t}(M_{\rm i})$ denotes the luminosity along the isochrone of age $t$,
and $\phi(M_{\rm i})$ is the initial mass function (IMF).  This latter
is usually normalized so that the total SSP mass is equal to a known
quantity $M_{\rm T}$, e.g.\ $\int_{0}^{\infty} M_{\rm i}\,\phi(M_{\rm
i})\, \diff M_{\rm i} = M_{\rm T} $.  Integrated magnitudes and
colours follow straightforwardly from the quantities $L_{\lambda}^{\rm
SSP}(t)$.

From the above equation, and taking a number of justified
approximations (see Girardi \& Bertelli 1998), we can also recover the
so-called ``{\it fuel consumption theorem}'' of Renzini \& Buzzoni
(1986):
        \beq
L_{\rm bol}^{\rm SSP}(t) = L_{\rm bol}^{\rm MS}(t) +
        A_{\rm H} \, b(t) \, \sum_{j} F_j(M_{\rm TO}) \,,
        \label{eq_fct}
        \eeq
In brief, this equation tells us that the integrated bolometric
luminosity of a SSP of given age can be expressed as the sum of the
integrated luminosity along the main sequence (MS), $L_{\rm bol}^{\rm
MS}(t)$, and a quantity that is proportional to the {\it total fuel},
$F_{\rm T} = \sum_{j} F_j(M_{\rm TO})$, consumed during all the
post-main sequence phases of a star with mass equal to the turn-off
mass, $M_{\rm TO}$.  In this equation $b(t)$ is the evolutionary flux,
i.e. the number of stars leaving the MS per unit time.  The fuel
consumption, $F_j$, represents the bolometric light emitted by all
stars in the post-MS evolutionary stage $j$. However, instead of being
expressed in energy units, this light is expressed in units of
``equivalent-mass of hydrogen burnt'' (e.g. in \Msun).  This
energy-to-mass conversion is properly taken into account by the
constant $A_{\rm H}$ in Eq.~(\ref{eq_fct}), which expresses the
efficiency (i.e. energy released per unit mass) of H-burning
reactions.  In this work we adopt $A_{\rm H} =9.75\times10^{10}\,
L_{\odot}\, {\rm yr}\, \Msolar^{-1}$, which is derived from the net
$Q$-value of H-burning reactions via the CNO cycle.

Since the burning of a given mass of helium provides only one tenth of
the energy that is provided by the nuclear consumption of the same
mass of hydrogen (i.e. $A_{\rm He} \sim 0.1\, A_{\rm H}$), $F_j(M_{\rm
TO})$ is then given by
        \beq
F_j(M_{\rm TO}) \simeq \Delta M\sub{H}_j + 0.1\,\Delta M\sub{He}_j \;,
        \label{eq_fuel}
        \eeq 
where $\Delta M\sub{H}_j$ and $\Delta M\sub{He}_j$ are the masses of H
and He, respectively, nuclearly burnt during the $j$-th post-main 
sequence evolutionary phase of the star of initial mass \Mto.

From Eqs.~(\ref{eq_isochrone}) and (\ref{eq_fct}) above, it is also
clear that
	\beq
F_j(M_{\rm TO}) = \frac{1}{A_{\rm H}\,b(t)} \int_{j} 
	\phi(M_{\rm i})\, L_{t}(M_{\rm i})\, 
	\diff M_{\rm i}     \,,
        \label{eq_light}
	\eeq
where the integration is limited to the section of the isochrone with
age $t$ that corresponds to the $j^{\rm th}$-evolutionary stage only.

On the other hand, if we are dealing with stellar evolutionary tracks
instead of isochrones, the fuel of a star with initial mass $M_{\rm
i}$ can be calculated as the time-integral of the luminosity over the
duration of that phase, divided by the constant $A_{\rm H}$:
	\begin{equation}
F_j(M_{\rm i}) = \frac{1}{A_{\rm H}}\int_j L_{M_{\rm i}}(t) \,
		{\rm d}t \,.
	\label{eq_flum}
	\end{equation}
In the previous equations, $F_j$ is expressed in $M_{\odot}$, $L$ is
given in $L_{\odot}$, and the independent variable $t$ is expressed in
yr.

From the aforementioned definitions it is immediately clear that the
$F_j$ are important quantities, because they relate basic properties
of stellar structures (Eq.~\ref{eq_fuel}) to the integrated light of
SSPs (Eq.~\ref{eq_light}).  Examples of the practical application of
the above concepts can be found in e.g.\ Greggio \& Renzini (1990)
and Renzini (1998).  It is just the stellar fuel
consumption and its main features the main subject of the analysis we
will develop in the following.

\section{Computing the fuel consumption}
\label{sec_fuel}
Let us now concentrate on the problem of determining, in a generalised
fashion, the quantities $\Delta M\sub{H}_j$ and $\Delta M\sub{He}_j$,
or equivalently $F_j$, corresponding to a stellar model of given
initial mass $M_{\rm i}$.
 
We can devise two methods respectively based on
	\begin{enumerate}
\item ``exact'' numerical integration along the track
\item analytical approximations 
	\end{enumerate}  

With the first approach the fuel consumed during the $j^{\rm
th}$-evolutionary phase can be calculated from the stellar track by
simply using Eq.~(\ref{eq_flum}), or, alternatively, by integrating
along the corresponding section of the isochrone with $M_{\rm TO} =
M_{\rm i}$ as in Eq.~(\ref{eq_light}).

The second method relies on simple approximate relations, involving
essentially the size of the fuel-exhausted core, and the
increase/decrease in the mean mass-coordinate of the overlying burning
shell.  In the context of this approach, we will present here an
analytical formalism to derive the fuel, developed in detail for low-
and intermediate-mass stars, and briefly sketched for massive stars.
The relations are of easy practical application as they contain
quantities that can be extracted from available stellar models.

If the first approach is to be preferred in terms of accuracy -- as it
straightforwardly follows from the definition of fuel consumption
itself (see Sect.~\ref{sec_ssp}) --, the second approach has the merit
of expliciting intimate inter-dependences between (apparently)
different properties of stellar populations (i.e. light emission and
chemical enrichment).  The accuracy of the formalism will be tested by
comparison to the results obtained with the ``exact'' method.
    
\begin{figure*}
\resizebox{\hsize}{!}{\includegraphics{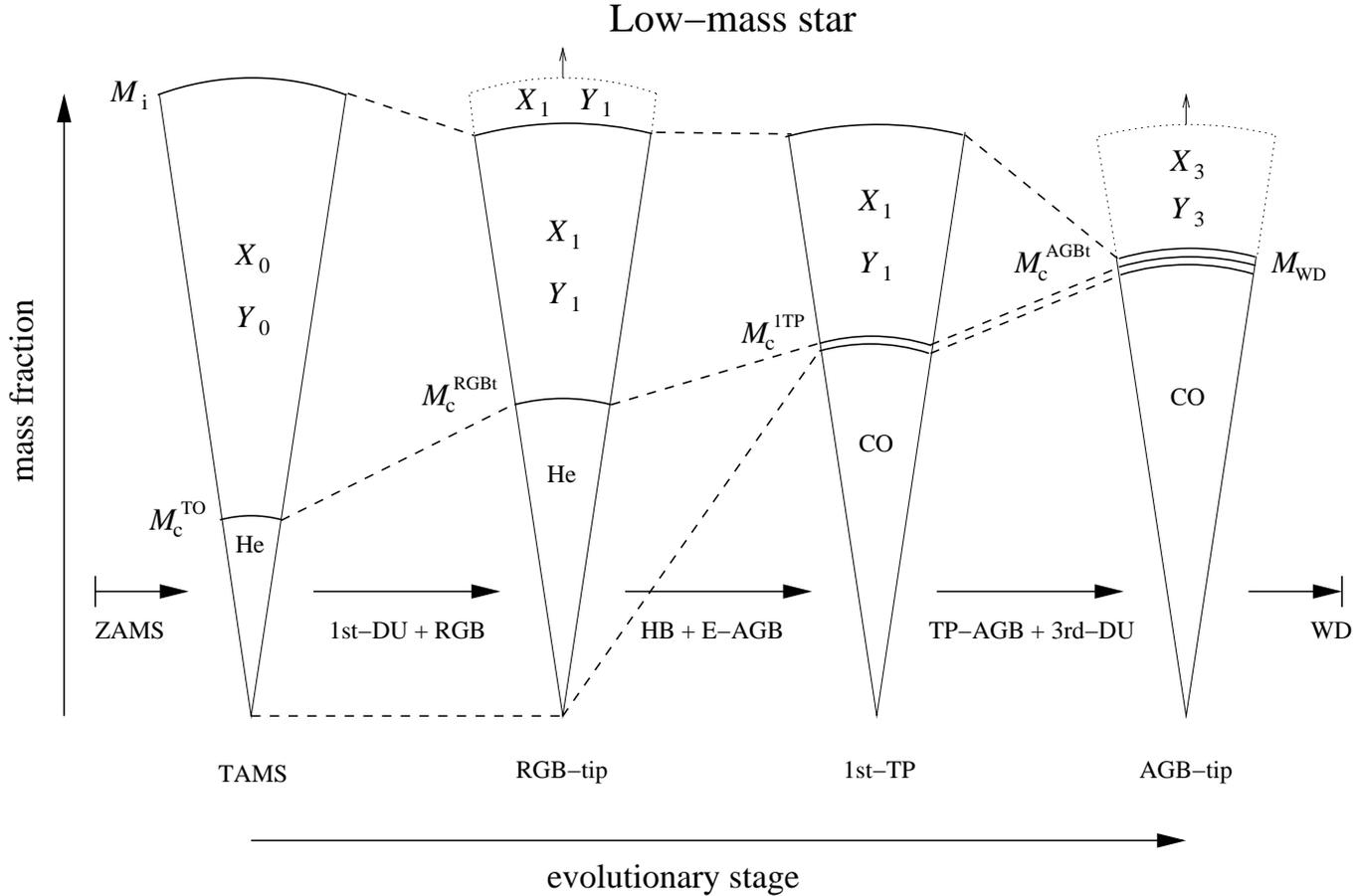}}
\caption{Schematic representation of the structure of a low-mass stars
at several evolutionary stages, namely (from left to right): at the
main sequence termination (TAMS), at the RGB tip, at the onset of
thermal pulses on the AGB (1st-TP), and at the last thermal pulse
(AGB-tip). These phases comprehend the whole post-main sequence
nuclear life of these stars.  The bold solid lines (connected by
dashed lines between stages), delimit the following boundaries (from
top to bottom) between: the stellar surface and the mass lost by
stellar winds, the H-rich envelope and the (H-exhausted) He core, and
the He and CO cores.  The chemical composition $(X,Y)$ of the H-rich
envelope is modified by the dredge-up (DU) episodes as marked. During
the RGB and AGB phases, mass-loss significantly reduces the stellar
mass (regions delimited by dots).  See text for a complete
discussion.}
\label{fig_low}
\end{figure*}

\begin{figure*}
\resizebox{\hsize}{!}{\includegraphics{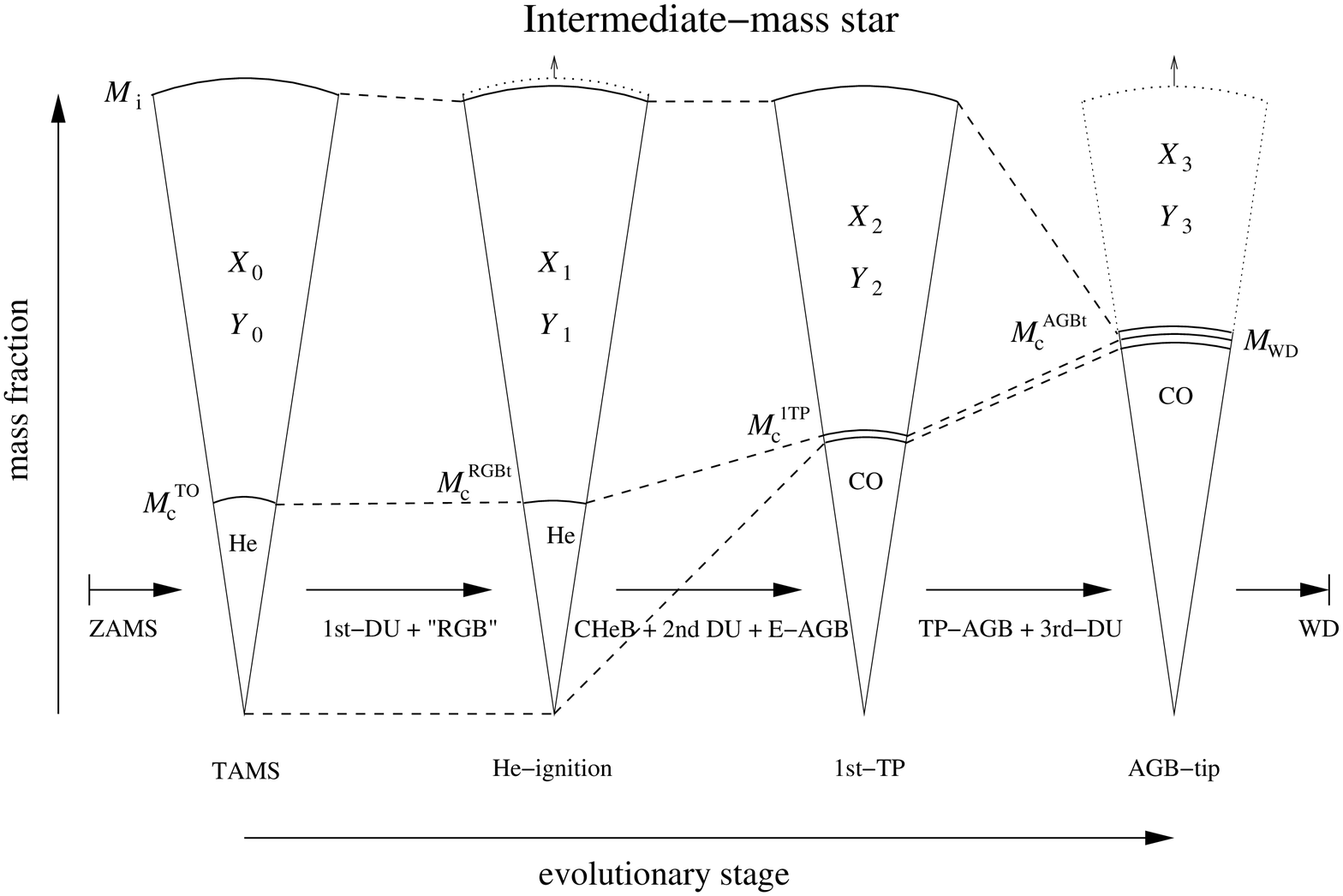}}
\caption{The same as Fig.~\protect\ref{fig_low} but for an
intermediate-mass star. In this case, the RGB phase is almost absent
and with negligible mass loss.}
\label{fig_int}
\end{figure*}

\subsection{Basic considerations}
\label{sec_basic}

Two fundamental ingredients in the calculation of the fuel are the
mass-coordinates of the H- and He-exhausted cores. This is already
suggested by Eq.~(\ref{eq_fuel}), that expresses the fuel in terms of
two contributions -- $\Delta M_{\rm H}$ and $\Delta M_{\rm He}$ --
which are obviously related to the growth of the H- and He-exhausted
cores, respectively.  In this sense we may say that, as the star
evolves, the He and CO cores keep track of the fuel being consumed, by
accreting the ashes of the nuclear burnings occurring above them.  We
may also advance that, in general, the derivation of the fuel will
involve the difference between the core masses at some ``initial'' and
``final'' stages of the phase under consideration.

From Eq.~(\ref{eq_fuel}), we can also derive some general rules that
are valid for any increase of the core mass $\Delta M_{\rm c}$. If we
are considering a H-exhausted core that has grown in mass by simply
transforming H into $^{4}$He, the corresponding fuel is increased by
$\Delta F_j=X\, \Delta M_{\rm c}$, where $X$ is the original abundance
of H (in mass fraction) in the material nuclearly burnt. If instead,
we are dealing with a He-exhausted core that accretes the products of
He-burning (mainly carbon and oxygen via the $\alpha$-capture
reactions), we would have $\Delta F_j=0.1\, \Delta M_{\rm c}$.
Finally, if during the same evolutionary stage, a core has grown in
mass through the successive conversions of H into He and then He into
CO, we would have $\Delta F_j=(X+0.1)\, \Delta M_{\rm c}$.

However, besides these general and quite simple considerations we
should add others which are equally important, though (probably)
somewhat less intuitive.  We remark, in fact, that the knowledge of
the $M_{\rm c}$ is not sufficient to estimate the fuel, since part of
the burnt material initially deposited in the core may be reduced by
{\sl dredge-up} events.  The net effect is that products of nuclear
burnings previously occurred {\sl in situ} in core regions (burning
shells included) are carried away to the outermost layers (dredge-up),
and a fraction of them can also be irreversibly lost by the star (mass
loss).  Then, we can already expect that this bit of information
subtracted from the core is intimately related to the {\sl stellar
yields}, as we will better explicit in Sect.~\ref{sec_analytic}.

To allow an easier understanding of the formalism developed in this
study let us recall now a few relevant aspects of dredge-up and
mass-loss processes characterising the evolution of low- and
intermediate-mass stars.

The dredge-up episodes -- caused by the penetration of the convective
envelope into regions previously affected by nucleosynthetic processes
-- typically occur in red-giant stages (i.e. RGB and AGB) between
major phases of core nuclear burning, possibly producing two effects:
	\begin{itemize}
\item changes in the chemical composition of the envelope, and 
\item inward shift of (either one or both) the hydrogen-helium and
helium-CO discontinuities (or chemical profiles).
	\end{itemize}
  
Evolutionary stellar calculations (e.g. Girardi et al. 2000; Marigo et
al. 1999) predict that the $1^{\rm st}$ dredge-up occurs at the base
of the RGB in low- and intermediate-mass stars of any mass, the
$2^{\rm nd}$ dredge-up takes place only in stars with initial mass
$M_{\rm i} \ga 3 - 4\, M_{\odot}$ at the beginning of the Early-AGB
(E-AGB), and the $3^{\rm rd}$ dredge-up\footnote{this term comprehends
all recurrent dredge-up events during the TP-AGB phase} may occur at
thermal pulses during the TP-AGB phase.

The $1^{\rm st}$ and $2^{\rm nd}$ dredge-up bring up to the surface
material which has undergone H-burning, with a consequent net
enrichment of $^{4}$He in the envelope. The third dredge-up can
contribute to increase the surface $^{4}$He abundance as well, at the
same time injecting into the envelope products synthesized at He-shell
flashes, mainly $^{12}$C and $^{16}$O of {\sl primary synthesis}.

In addition to the aforementioned dredge-up episodes, the surface
chemical composition of TP-AGB stars with $M_{\rm i} \ga 3.5
M_{\odot}$ can be altered by the occurrence of hydrogen-burning via
the CNO-cycle at the base of their convective envelopes (hot-bottom
burning; HBB).  This latter process also increases the surface
abundance of $^{4}$He, and that of $^{14}$N at the expenses of the
newly dredged-up $^{12}$C (and possibly $^{16}$O).

Another crucial process characterising the evolution of low- and
intermediate-mass stars is {\sl mass loss} via stellar winds, both
during the RGB (of low-mass stars) and AGB phases.  From both
theoretical and observational arguments we get indications that during
the final stages of the AGB evolution of these stars, mass loss is
able to almost completely strip off their mantles. The final result is
a white dwarf consisting almost completely of a C-O core, and with
very thin (in mass) superficial layers of H and He.

It is just the combination of surface composition changes (due e.g. to
dredge-up events) and mass loss that determines the contribution of
these stars to the chemical enrichment of the interstellar medium
(ISM). Such contribution is commonly quantified through the {\sl
stellar yields}, defined as the amount of newly synthesized elements
ejected into the ISM.

\subsection{Analytical derivation}
\label{sec_analytic}

In order to guide our discussion, Figs.~\ref{fig_low} and
\ref{fig_int} schematically present the basic structure of low- and
intermediate-mass stars at some key-evolutionary stages.  In our
scheme, each star starts as a chemically homogeneous zero-age main
sequence (ZAMS) configuration, and ends up as a CO white dwarf with
virtually no H- or He-rich envelopes. Thus, the evolutionary stages
depicted comprehend its whole post-MS nuclear evolution.

Each stage  
is identified by the corresponding mass of the H-exhausted
core $M_{\rm c}$, namely:
	\begin{itemize}
	\item the termination of the main sequence (TAMS).
	This is equivalent to the
	turn-off (TO) point of the corresponding isochrone (or SSP) 
	with $M_{\rm c}^{\rm TO}$;
	\item the tip of the RGB (or the stage of central He ignition)
with $M_{\rm c}^{\rm RGBt}$;
	\item the first thermal pulse on the AGB with $M_{\rm c}^{\rm
1TP}$; and
	\item the tip of the AGB with $M_{\rm c}^{\rm AGBt}$ (notice
that $M_{\rm c}^{\rm AGBt}$ coincides with the remnant mass $M_{\rm
WD}$).
	\end{itemize}
Moreover, it is worth noticing that during the stages under
consideration not only the mass-coordinate of the core(s), but also
its chemical composition (either a He-core, or a CO-core surrounded by
a very thin He-rich layer) can be easily singled out in stellar
models.  This can be appreciated looking at the lines depicted in
Figs.~\ref{fig_low} and \ref{fig_int} that schematically describe the
evolution of the H--He and He--CO discontinuities.

Outside the H-exhausted core, there is an envelope whose mass is
reduced by both the core growth and mass loss, but can be
temporarily increased (even if the former decrease is by far dominant)
by dredge-up events.  These latter events also change the envelope chemical
composition.  We denote by $X_j$, $Y_j$ -- with $j=0,1,2,3$ -- the
envelope fractional abundances (by mass) of hydrogen and helium
respectively, where the superscripts indicate the original abundances
if $j=0$, and those after the $j^{\rm th}$-dredge-up event if $j > 0$.

In general, with the term ``core mass'' we denote the mass-coordinate
of a specified chemical discontinuity (e.g. the mass of the
H-exhausted core corresponds to the H-He discontinuity). This
definition, however, cannot always be applied.  For instance, stellar
models at the TAMS do not show any sharp H--He discontinuity, but
rather a chemical profile -- left by either the radiative core or the
recession of the convective core --, so that $\Mcore^{\rm TO}$ does
not correspond to any well-defined mass-coordinate inside the star.
Anyhow, we can find a physically equivalent definition for
$\Mcore^{\rm TO}$, that is
	\beq
	\Mcore^{\rm TO} = \frac{1}{X_0}\, \int_{0}^{M_{\rm i}} 
                          [X_0 - X(M_r)]\, {\rm d}M_r \, ,
	\eeq
where $X(M_r)$ is the H abundance at the mass-coordinate $M_r$.  In
practice, $\Mcore^{\rm TO}$ corresponds to the mean value of $M_r$
across the chemical profile, weighted by the difference between the
initial and local H abundance.  We can also notice that the total mass
of H burnt during the MS is equal to $X_0\, \Mcore^{\rm TO}$.
Moreover, Eq.~(\ref{eq_flum}) offers an alternative way to compute
$\Mcore^{\rm TO}$ from evolutionary tracks:
	\beq
\Mcore^{\rm TO} = \frac{1}{X_0\, A_{\rm H}} 
	\int_{\rm MS} L_{M_{\rm i}}(t) \, {\rm d} t \,\,\,.
	\label{eq_mcto}
	\eeq 

At this point we have defined (and illustrated in Figs.~\ref{fig_low}
and \ref{fig_int}) all the ingredients necessary to estimate the
nuclear fuel consumed in different evolutionary stages of low- and
intermediate-mass stars.

\subsubsection{The total fuel}

Hereinafter we will briefly refer to total fuel as that consumed
during the entire post-main sequence evolution.  This can be evaluated
with
        \beq 
F_{\rm T} \simeq (X_0 + 0.1) \, \Mcore^{\rm AGBt} - X_0\, \Mcore^{\rm TO} + 
	F_{\rm y}^{\rm He} + F_{\rm y}^{\rm CO} \, ,
        \label{eq_totfuel}
        \eeq
where $(X_0 + 0.1) \, \Mcore^{\rm AGBt} - X_0\, \Mcore^{\rm TO}$
represents the fuel required to convert a He-core of mass $\Mcore^{\rm
TO}$ into a CO-core of mass $\Mcore^{\rm AGBt}$ (see
Figs.~\ref{fig_low} and \ref{fig_int}); $F_{\rm y}^{\rm He}$ and
$F_{\rm y}^{\rm CO}$ represent the contributions to the fuel related
to the stellar yields of $^{4}$He, and $^{12}$C $+$ $^{16}$O elements
(produced by He-burning reactions), respectively. In this paper,
$M_{\rm y}(i)$ \footnote{For all the definitions related to yields, we
refer to the classical work by Tinsley (1980), and to the recent
calculations presented by Marigo (2001).} denotes the stellar yield of
element $i$ (in mass units).

The term referred to $^{4}$He is, by definition, simply 
	\beq
F_{\rm y}^{\rm He} = M_{\rm y}(\rm He) \, ,
	\eeq
with
	\beq
 M_{\rm y}({\rm He}) = \int_0^{\tau^*} \left[ Y(t)-Y_0 \right] 
	\frac{{\rm d}M}{{\rm d}t} {\rm d}t
	\label{eq_heyield}
	\eeq
where the integral is carried out over the entire stellar lifetime
$\tau^*$; $Y(t)$ and ${\rm d}M/{\rm d}t$ are the current (at time $t$)
surface $^{4}$He abundance and mass-loss rate, respectively.

The newly synthesised CO nuclei from He-burning reactions are clearly
of {\it primary origin}.  This applies, for instance, to $^{12}$C and
$^{16}$O nuclei brought up to the surface by the third dredge-up, as
they are synthesised at He-shell flashes during the TP-AGB evolution.
Denoting with $M_{\rm y}^{\rm P}({\rm CO})$ the corresponding primary
yield (see Eq.~\ref{eq_ycoagb} below), we can use the approximation:
	\beqa
F_{\rm y}^{\rm CO} & = & 1.1\,(1-Y') \,M_{\rm y}^{\rm P}({\rm CO}) +
	0.1 \, Y' \, M_{\rm y}^{\rm P}({\rm CO})  \\ 
	\label{eq_fco}
	 & = & (1.1 - Y') \, M_{\rm y}^{\rm P}({\rm CO}),  
	\nonumber  
	\eeqa
where 
	\beq
Y' = \frac{Y_{1,2}}{X_{1,2}+Y_{1,2}} \;\: .
	\eeq
This latter quantity involves H and $^{4}$He abundances at the onset
of the TP-AGB phase, i.e. after the $2^{\rm nd}$ dredge-up or, if this
latter does not occur as in lower mass stars, after the $1^{\rm st}$
dredge-up.  Equation~\ref{eq_fco} accounts for the fact that a
fraction $Y'$ of $M_{\rm y}^{\rm P}({\rm CO})$ derives directly from
nuclear burning of original helium, whereas the complementary fraction
$1-Y'$ is synthesized starting from original hydrogen through the
sequence of both H- and He-burning.

It is worth remarking that all the quantities that enter in the
evaluation of Eq.~(\ref{eq_totfuel}) -- i.e. the core masses $M_{\rm
c}^{\rm TO}$, the remnant mass $M_{\rm c}^{\rm AGBt} = M_{\rm WD}$,
the chemical abundances after dredge-up episodes, $X_j$ and $Y_j$, and
the stellar yields $M_{\rm y}({\rm He})$ and $M_{\rm y}^{\rm P}({\rm
CO})$ -- can either be easily derived from published stellar tracks,
or are already tabulated in papers that deal with chemical yields.

Finally, the two components of Eq.~(\ref{eq_fuel}) can be also
distinguished:
	\beqa
\Delta M_{\rm H} & \simeq & X_0 \, 
	(M_{\rm c}^{\rm AGBt} - M_{\rm c}^{\rm TO}) + M_{\rm y}({\rm He}) \\
	\nonumber
                 &           & + (1-Y') \, M_{\rm y}^{\rm P}({\rm CO}) 
	\label{eq_mhdup}
	\eeqa
	\beq
\Delta M_{\rm He} \simeq M_{\rm c}^{\rm AGBt} + M_{\rm y}^{\rm P}({\rm CO}) 
	\label{eq_mhedup}
	\eeq

The above equations can be slightly modified when we consider stars
(with $M_{\rm i} \ga 3.5 M_{\odot}$) experiencing hot-bottom burning
(in addition to the dredge-up episodes) during the TP-AGB phase.  In
this case the dredged-up $^{12}$C (and possibly $^{16}$O) may be
converted into $^{14}$N, so that it is advisable to replace $M_{\rm
y}^{\rm P}(\rm CO)$ in Eqs.~(\ref{eq_fco}) -- (\ref{eq_mhedup}) with
the total {\em primary} yield of the CNO elements, $M_{\rm y}^{\rm
P}(\rm CNO)$.

With the aid of Figs.~\ref{fig_low} and \ref{fig_int}, we can easily
specify the contributions to the total fuel coming from the different
post-main sequence shell(s)-burning phases:
	\beq
F_{\rm T} = F_{\rm RGB} + F_{\rm CHeB+EAGB} + F_{\rm TP-AGB}
	\eeq
where the rigt-hand side terms are those derived in the following.

\subsubsection{The RGB fuel}
\label{sssec_rgb}

        \beqa 
        \label{eq_rgbfuel}
F_{\rm RGB} & \simeq & (X_1 + 0.1) \, (\Mcore^{\rm RGBt} - 
	\Mcore^{\rm TO})\\
 & &	+ (Y_1 - Y_0) \, (M_{\rm i} - \Mcore^{\rm RGBt})\, .
	\nonumber
        \eeqa
Here, the term related to the $1^{\rm st}$ dredge-up is given by the
amount of newly synthesized helium mixed up into the envelope mass
$M_{\rm i} - \Mcore^{\rm RGBt}$, where $M_{\rm i}$ is the initial
stellar mass.  We can notice that in this case the term $F_{\rm
y}^{\rm CO}=0$, as the first dredge-up involves material that has
experienced only H-burning reactions.

This equation requires the evaluation of $\Mcore^{\rm RGBt}$, the core
mass at He-ignition. This quantity is normally tabulated only for
low-mass stars, because it critically determines the luminosity of HB
and red clump stars. Moreover, in low-mass stars we always have
$\Mcore^{\rm RGBt}-\Mcore^{\rm TO} > 0$, since the mass of the
electron-degenerate core must grow up to $0.45 - 0.50\, M_\odot$
before He can ignite.

For intermediate-mass stars, the RGB phase is practically missing.
Anyhow, for the sake of a uniform notation, we can still refer to
$F_{\rm RGB}$ as the fuel (very little) consumed from the TAMS up to
central He-ignition.  Moreover, we notice that, in general, the
quantity $\Mcore^{\rm RGBt}$ is not easily singled out in stellar
tracks.  In fact, in intermediate-mass models the H-burning shell is
rather thick (in mass), so that a sharp H--He discontinuity cannot be
recognised.  To overcome this difficulty we may adopt a simple
approach, that is to derive $\Mcore^{\rm RGBt}$ from:
	\beq
\Mcore^{\rm RGBt} \simeq \Mcore^{\rm TO} -  M_{\rm dred} \,  ,
	\eeq
where
	\beq
M_{\rm dred} \simeq \frac{Y_1-Y_0}{1-Y_1}\, (M_{\rm i}-\Mcore^{\rm TO})\, . 
	\eeq
In other words we assume that during the ``RGB'' phase the core is
only affected by the $1^{\rm st}$ dredge-up, which reduces its mass
(by the amount $M_{\rm dred} = \Mcore^{\rm TO}-\Mcore^{\rm RGBt}$),
and increases the envelope helium content from $Y_0$ to $Y_1$
(normally a tabulated quantity).  Actually the $1^{\rm st}$ dredge-up
usually proceeds inward across a chemical profile, instead of a H-He
discontinuity implicitly assumed in the latter formula.  Therefore,
the above equations give just a crude approximation of $\Mcore^{\rm
RGBt}$ for intermediate-mass stars.

\subsubsection{The core He-burning + early-AGB fuel}

        \beqa 
F_{\rm CHeB+EAGB} & \simeq & (X_1 + 0.1) \, \Mcore^{\rm 1TP} - 
	X_1\,\Mcore^{\rm RGBt} \\
 & &	+ (Y_2 - Y_1) \, (M_{\rm i} - \Mcore^{\rm 1TP})\, .
	\nonumber
        \label{eq_chebfuel}
        \eeqa
This expression can be easily understood when compared to
Eq.~(\ref{eq_rgbfuel}).  The term related to the $2^{\rm nd}$
dredge-up is zero whenever this does not occur (i.e. for $M \la
3-4$~\Msun).
 
$\Mcore^{\rm 1TP}$ is normally a tabulated quantity, as in e.g. Marigo
et al.\ (1998).

Since core-He burning tends to convert the whole He-core into a
CO-core, and since during most of the E-AGB phase the H-burning shell
is off, one can also give a rough estimate to the fuel that comes from
the E-AGB phase only:
        \beqa 
F_{\rm EAGB} & \sim & 0.1 \, (\Mcore^{\rm 1TP} - 
	\Mcore^{\rm RGBt})\\
 & &	+ (Y_2 - Y_1) \, (M_{\rm i} - \Mcore^{\rm 1TP})\, .
	\nonumber
        \label{eq_eagbfuel}
        \eeqa

Finally, we remind the reader to consider the cautionary remarks,
expressed in Sect.~\ref{sssec_rgb}, on the definition of $\Mcore^{\rm
RGBt}$ in the case of intermediate-mass stars.

\subsubsection{The TP-AGB fuel}

        \beqa 
        \label{eq_agbfuel}
F_{\rm TP-AGB}  & \simeq & (X_{1,2} + 0.1)\, (\Mcore^{\rm AGBt} - 
	\Mcore^{\rm 1TP})  \\ 
	\nonumber
  & &  + M^{\rm TP-AGB}_{\rm y}{\rm (He)} \\
	\nonumber
  & &  + (1.1 - Y')\, M^{\rm P}_{\rm y}({\rm CO})\, .
	\nonumber
        \eeqa 

It should be noticed that the quantity related to the stellar yield of
helium, $M^{\rm TP-AGB}_{\rm y}{\rm (He)}$, should be scaled with
respect to the elemental abundance at the beginning of the TP-AGB
phase, and not to the initial one as in the standard definition of
stellar yields (and in Eq.~\ref{eq_heyield}).  Then, the $^{4}$He
contribution can be expressed as
	\beq
M^{\rm TP-AGB}_{\rm y}{\rm (He)} = 
	\sum_{j=1}^{N_{\rm p}} \left[ 
	(Y_{j+2}- Y_{\rm 1,2}) \, \Delta M_j \right] \,,
	\label{eq_yheagb}
	\eeq
where $Y_{j+2}$ corresponds to the abundance after the $j^{\rm th}$
thermal pulse (or $(j+2)^{\rm th}$-dredge-up event), and $\Delta M_j$
denotes the mass ejected during the $j^{\rm th}$ pulse-cycle.  The
summation is performed over the total number, $N_{\rm p}$, of pulse
cycles.  A functional form analogous to Eq.~(\ref{eq_yheagb}) should
apply to $M^{\rm P}_{\rm y}({\rm CO})$ as well, i.e.:
	\beq
M^{\rm P}_{\rm y}{\rm (CO)} = 
	\sum_{j=1}^{N_{\rm p}} \left[ 
	X^{\rm P}({\rm C}+{\rm O})_{j+2} 
	\,\, \Delta M_j \right] \, ,
	\label{eq_ycoagb}
	\eeq
where $X^{\rm P}({\rm C}+{\rm O})_{j+2}$ denotes the primary carbon
and oxygen abundance in the envelope after the $j^{\rm th}$ thermal
pulse. It should be noticed that, in this case, the surface chemical
composition at the onset of the TP-AGB phase does not contain any
element of primary origin, i.e. the scaling term (corresponding to
$Y_{1,2}$ in Eq.~\ref{eq_yheagb}) is set to zero in
Eq.~(\ref{eq_ycoagb}).
 
Finally, we recall that in case of hot-bottom burning a better
approximation is achieved using $M^{\rm P}_{\rm y}({\rm CNO})$ instead
of $M^{\rm P}_{\rm y}({\rm CO})$ (see earlier in this section).

\subsubsection{Further corrections} 

The analytical relations presented so far consider the main
contributions to the fuel, (i.e. nuclear burnings), but neglect other
terms that possibly take part to the energy balance of a star, such as
neutrino losses, and energy gains due to gravitational contraction.
For the sake of completeness, these terms should be added to the
right-hand side of Eq.~(\ref{eq_totfuel}).  However, neglecting them
does not introduce a significant error in the evaluation of the fuel,
as these terms only determine small corrections not exceeding a small
percentage in most cases.

Anyhow, we can easily get a rough estimate of the gravitational energy
released by gravitational contraction during the evolution of low- and
intermediate-mass stars with
	\beq
\Delta E_{\rm G}  = - \frac{G \, M^2_{\rm WD}}{R_{\rm WD}}
	\eeq
which gives the order of magnitude of the gravitational energy of a
white dwarf with a mass $M_{\rm WD}$ and radius $R_{\rm WD}$. This
latter can be derived, for instance, adopting the mass-radius relation
for white dwarfs (as derived from Tuchman et al. (1983) basing on
Chandrasekhar (1939)):
	\beq
R_{\rm WD} = 0.019 \, (1 - 0.58\, M_{\rm WD})
	\eeq
(expressed here in solar units) and setting $M_{\rm WD} = \Mcore^{\rm
AGBt}$, i.e. the mass of the core left at the end of the AGB.

Then, the corresponding contribution to the stellar fuel (in mass
units) can be evaluated with
	\beq
\Delta F_{\rm G} = - \frac{1}{2}
	\frac{\Delta E_{\rm G}}{A_{\rm H}}
	\label{eq_fg}
	\eeq
where the factor $1/2$ accounts for the fact that only half of the
gravitational energy goes in radiation (Virial Theorem).  A simple
evaluation reveals that $\Delta F_{\rm G}$ is of order of just a few
hundredths of $M_\odot$.  The relative contribution of this term to
the total fuel will be shown in a few examples discussed in
Sect.~\ref{sec_examp}.

As far as neutrino losses are concerned, in the evolutionary phases
under consideration they can be essentially of two kinds:
	\begin{enumerate}
	\item
Neutrino losses via $\beta$ decays: they carry away a fraction of the
energy produced by nuclear reactions. Since the constant $A_{\rm H}$
(Eq.~\ref{eq_fct}) includes only the energy that is effectively
available to the star -- thus excluding neutrinos -- no further
correction is needed in our equations.
	\item
Plasma neutrino losses: they cool the stellar cores during the RGB and
AGB phases. However, the energy they carry away is always very small
if compared to the stellar surface luminosity. In RGB stars, just some
thousandths of the stellar energy leaves the star in the form of
plasma neutrinos, whereas in the AGB this fraction is of a few
hundredths-- reaching a maximum of about 0.06 in the most luminous and
massive AGB stars. Thus, they would lead to very small corrections --
a few percent in extreme cases -- to the fuel, that we will neglect in
our approach.
	\end{enumerate}

\subsection{The case of massive stars}
\label{sec_massive}

Massive stars evolve through successive nuclear burning stages in
their interior up to the supernova (SN) explosion, possibly leaving
either a neutron star or a black hole as remnant. In comparison to the
case of low- and intermediate-mass stars, several factors hamper the
derivation of simple and accurate formulas for the fuel consumed by
massive stars, including the complex succession of nuclear burnings in
the latest stages of hydrostatic core evolution, and the complex
energetics of the explosion event.

Nonetheless, we may still provide simple formulas in order to give
reasonable estimates to the fuel. First, we can suitably separate the
total fuel into two components:
	\beq
F_{\rm T} = F_{\rm pre-SN} + F_{\rm SN} 
	\eeq
that refer to the pre-SN, and SN contributions, respectively.  In
analogy to Eq.~(\ref{eq_totfuel}), the first term can be generalised
as:
        \beqa
        \label{eq_fuelpresn}
F_{\rm pre-SN} & \simeq & X_0\, (\Mcore^{\rm He}-\Mcore^{\rm CO}) \\ 
\nonumber
	& & \sum_{i \ge 2} (X_0 + x_{i,1}) \, 
		(\Mcore^{\rm i} - \Mcore^{i+1}) \\ 
\nonumber
	& & - X_0\, \Mcore^{\rm TO} 
	+ F_{\rm y}^{\rm He} + F_{\rm y}^{CO}  \, .
        \eeqa
The first term and the summation in the second term account for all
the contributions to the fuel involved in the formation of the
``onion-skin'' structure inside the H-exhausted core. Here the
increase of the index $i$ corresponds to more and more advanced
nuclear burnings (e.g. $i=1$ refers to H-burning, $i=2$ is related to
He-burning, etc.) and $\Mcore^{i}$ denotes the mass-coordinate of the
$i^{\rm th}-$element exhausted core (e.g. $\Mcore^{\rm He}$ for $i=1$,
$\Mcore^{\rm CO}$ for $i=2$, etc).  In general, the quantity $x_{i,j}$
is related to the amount of energy generated in the conversion of
nuclei $i$ into nuclei $j$, normalised to that provided by the
conversion of H into $^{4}$He.  It can be estimated from the
difference between the binding energies per nucleon, $B_A$, that
characterize the nuclei involved ($i =$ final, $j =$ initial):
	\beq
x_{i,j} = \frac {B_{A(i)} - B_{A(j)}} {B_{^4{\rm He}}}  
	\eeq
with respect to that of $^4{\rm He}$ (the main product of H-burning).
We can notice that, by definition, $B_{\rm H}=0$, and that $x_{2,1}
\simeq (B_{^{12}{\rm C}}-B_{^4{\rm He}})/B_{^4{\rm He}} \sim 0.1$, as
already seen in the previous sections.  In other words, $x_{i,1}$
represents the global efficiency -- relative to that of H-burning --
of all nuclear burnings (from that of He to the $i^{\rm th}$-one)
which have successively taken place in the shell of mass $(\Mcore^{i}
- \Mcore^{i+1})$.
                                           
Finally, the term $- X_0\, \Mcore^{\rm TO} + F_{\rm y}^{\rm He} +
F_{\rm y}^{CO}$ has the same meaning as in Eq.~(\ref{eq_totfuel}), but
with a difference: here the chemical yields refer only to those
produced during the pre-SN evolution, that is, to the so-called {\em
wind yields}.

The practical applicability of Eq.~(\ref{eq_fuelpresn}) is determined
by the fact that the quantities $\Mcore^{\rm He}$, $\Mcore^{\rm CO}$,
and the wind yields, are normally presented in published pre-SN models
(see e.g. Maeder 1992; Woosley \& Weaver 1995; Portinari et al. 1998).
Conversely, the interior ``onion-skin'' structure is generally not
explicitly given.  Anyway, we can use simple approximations to derive
lower and upper limits to the second right-hand side term of
Eq.~(\ref{eq_fuelpresn}). A lower limit is obtained neglecting any
nuclear conversion into elements heavier than CO nuclei ($x_{2,1} =
0.1$), so that the term simply reduces to $(X_0 + 0.1) \Mcore^{\rm
CO}$. An extreme upper limit corresponds to supposing that the
He-exhausted core has been wholly converted into the iron-group
elements.  In this case, we have $x_{i,1}=x(^{12}{\rm C} \rightarrow
^{56}{\rm Fe}) \simeq 0.16$ and the second right-hand side term
becomes $(X_0 + 0.26) \Mcore^{\rm CO}$.

Therefore, these simple estimates suggest that $F_{\rm pre-SN}$ should
be comprised within a rather narrow range, of width $\sim 0.16
\Mcore^{\rm CO}$. This narrowness is determined by the relatively low
amount of energy that is available from the conversion of CO into
heavier elements.

Computing $F_{\rm SN}$ can be far more difficult, because of the
complicated energetics involved in the SN explosion.  Anyway, a simple
order-of-magnitude estimate comes from the conversion of the typical
energy irradiated by SNe, $10^{51}$~erg, into the equivalent mass of
burnt H:
	\beq
F_{\rm SN} \sim 0.1\,  M_{\odot} \,\, .
	\eeq
Therefore, we would expect that usually $F_{\rm SN} \ll F_{\rm
pre-SN}$.  

\begin{figure*}
\resizebox{\hsize}{!}{\includegraphics{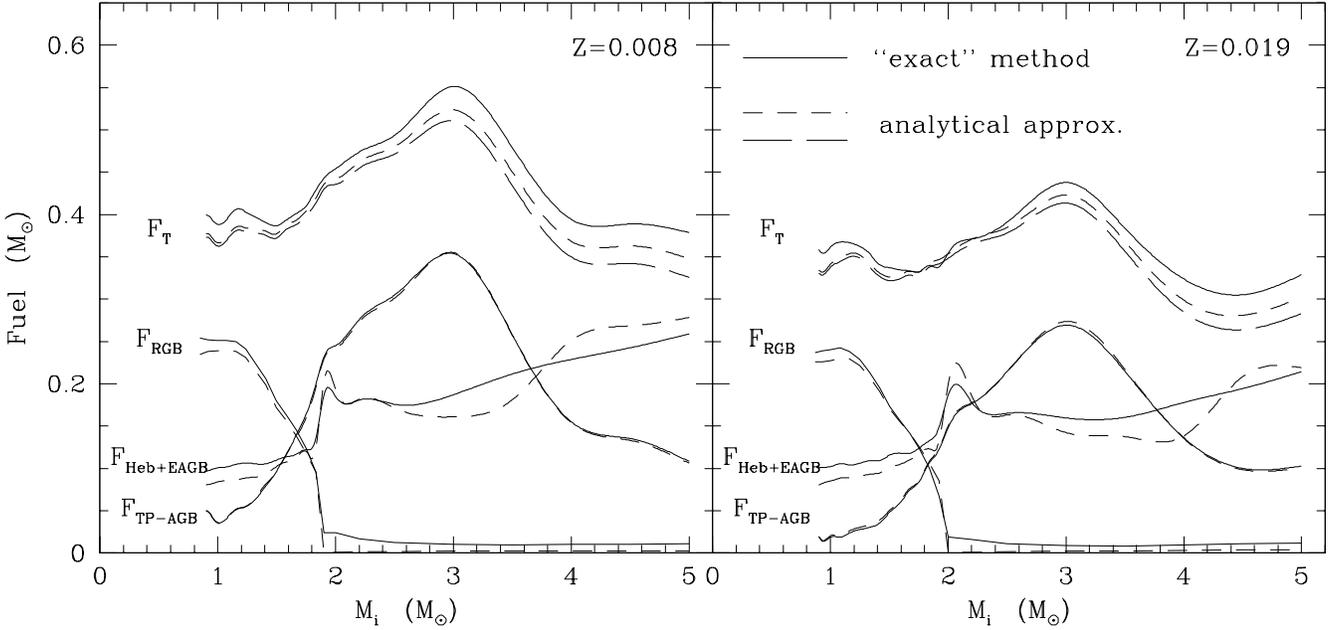}}
\caption{Accuracy checks.  Stellar fuel consumption as a function of
the stellar mass, calculated for two sets of stellar models with
different initial metallicity (as indicated) and considering various
post-MS phases (labelled nearby the corresponding curve).  The results
predicted by the ``exact'' method given by Eq.~\protect(\ref{eq_fuel})
is compared with those obtained by adopting the analytical formalism
here proposed (dashed lines). The total fuel is calculated both
without (long-dashed line) and with (short-dashed line) the
gravitational term expressed by Eq.~\protect(\ref{eq_fg}).}
\label{fig_3}
\end{figure*}

\section{Illustrative applications}
\label{sec_examp}
Here we will apply the analytical formalism derived in the
previous section to a few representative examples, in the domain of
low- and intermediate-mass stars. First, we will check the accuracy of
the formulas, applying them to a particular case in which we know in
advance that the stellar inputs are homogeneous and self-consistent
(Sect.~\ref{sec_accuracy}). For this specific case, we will analyse
the fraction of the fuel that is in the form of chemical yields
(Sect.~\ref{sec_chemeffect}). This will also allow us to discuss the
general features of the fuel as a function of the stellar mass and
metallicity, and to quantify the error we may introduce if we neglect
the effect of dredge-up episodes (and, in general, of chemical yields)
in estimates of the fuel consumption. Finally, we will illustrate a
case in which heterogeneous stellar data are used to model the emitted
light and chemical yields (Sect.~\ref{sec_theneed}). This will allow
us to remark on the consistency requirements for the input stellar data
in chemical-spectrophotometric models of stellar populations. 

\subsection{Accuracy checks}
\label{sec_accuracy}

The accuracy of the analytical formalism presented in
Sect.~\ref{sec_fuel} should be tested by comparing its predictions to
the corresponding results obtained with the ``exact'' method already
mentioned in Sect.~\ref{sec_ssp}, which is essentially based on the
calculation of the integral in Eq.~(\ref{eq_flum}).  Of course, this
can be done only if we have a {\em homogeneous set of stellar models},
providing tables with both (i) the evolutionary tracks (or
isochrones), and (ii) the chemical yields and core masses.

Here we adopt a set of low- and intermediate-mass evolutionary models
with $ 0.9 M_{\odot} \la M \le 5 M_{\odot}$, $Z_0=0.008$, $Y_0=0.25$
computed by Girardi et al. (2000) and Marigo et al. (1999), which
follow the evolution from the ZAMS up to the end of the AGB, and
predict the corresponding stellar yields (Marigo 2001). These models
are more extensively described in Appendix A. 

The outcome of the test is shown in Fig.~\ref{fig_3}, for two 
different initial metallicities ($Z_0=0.008$ and
$Z_0=0.019$).  It turns out that the analytical prescriptions
reproduce remarkably well the fuel calculated with the ``exact''
method, the relative difference ranging typically within $5\, \%$,
mostly concerning the RGB, TP-AGB, and the total post-MS
contributions.  As expected, the largest differences result in the
analytical derivation of $F_{\rm RGB}$ and $F_{\rm CHeB+EAGB}$ of
intermediate-mass stars, due to the difficulty of defining the core
mass-coordinate $\Mcore^{\rm RGBt}$ (see remarks in
Sect.~\ref{sssec_rgb}).

Figure~\ref{fig_3} displays also the effect of the correction term,
$F_{\rm G}$, calculated with Eq.~(\ref{eq_fg}), which is added to the
total fuel derived from Eq.~(\ref{eq_totfuel}). As expected, the
convergency towards the ``exact'' result is improved by a small amount
(few percentiles at maximum) with the inclusion of the gravitational
term.  Finally, we notice that even if some of the partial terms
(i.e. $F_{\rm RGB}$, $F_{\rm HeB+EAGB}$) may be not well reproduced,
the analytical prescription for the total post-MS fuel $F_{\rm T}$ is
quite accurate.  This reflects the small number of stellar parameters
involved in Eq.~(\ref{eq_totfuel}).

\subsection{The effect of surface chemical changes}
\label{sec_chemeffect}
          
\begin{figure}
\centering
\resizebox{0.8\hsize}{!}{\includegraphics{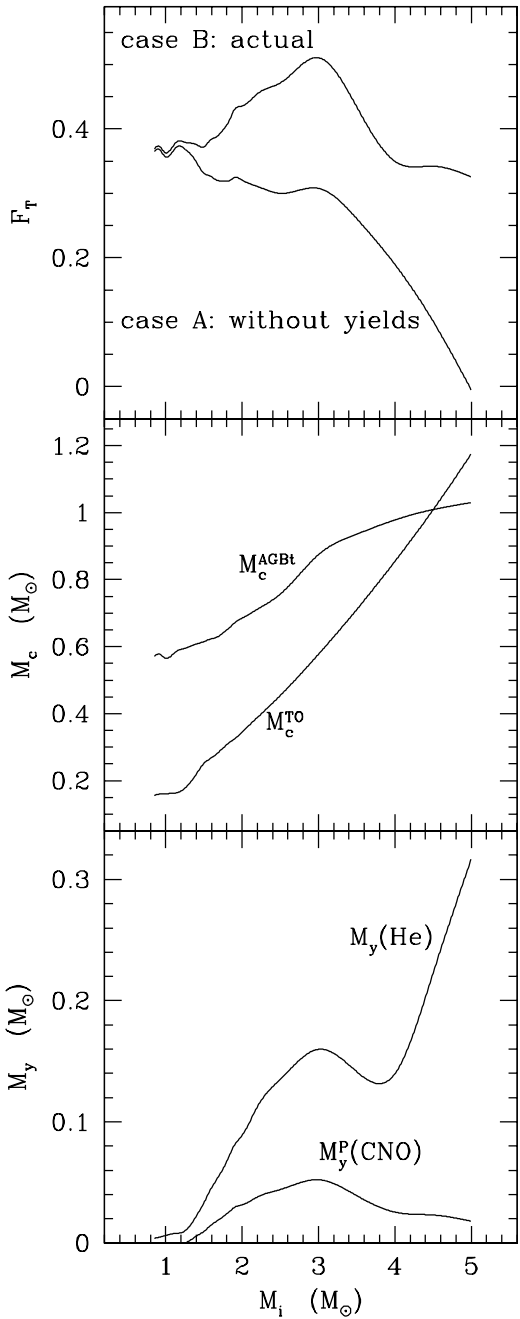}}
\caption{Upper panel: Total post-main sequence fuel consumption 
by low- and intermediate-mass models with initial composition
($Y_0=0.25$, $Z_0=0.008$), calculated 
with and without the  terms ($F_{\rm y}^{\rm He}$ and $F_{\rm y}^{\rm CO}$) 
related to the chemical yields
in Eq.~(\protect\ref{eq_totfuel}).
Middle panel: masses of the H-exhausted cores as defined in the text.
Bottom panel: chemical yields of $^{4}$He and primary CNO.
See text for more details}
\label{fig_1}
\end{figure}

Figure \ref{fig_1} displays the predicted stellar fuel consumed over
the post-main sequence evolution, as a function of the initial stellar
mass, again for the Girardi et al. (2000) and Marigo et al. (1999)
tracks.
For illustrative purposes, we first compute the fuel according to
Eq.~(\ref{eq_totfuel}) but neglecting the terms due to the dredge-ups
(i.e. setting $F_{\rm y}^{\rm He} = F_{\rm y}^{\rm CO} = 0$).  This is
referred to as case A.  The inclusion of these ``chemical'' terms
corresponds instead to case B, that is the correct derivation of the
fuel.

\begin{figure*}
\resizebox{\hsize}{!}{\includegraphics{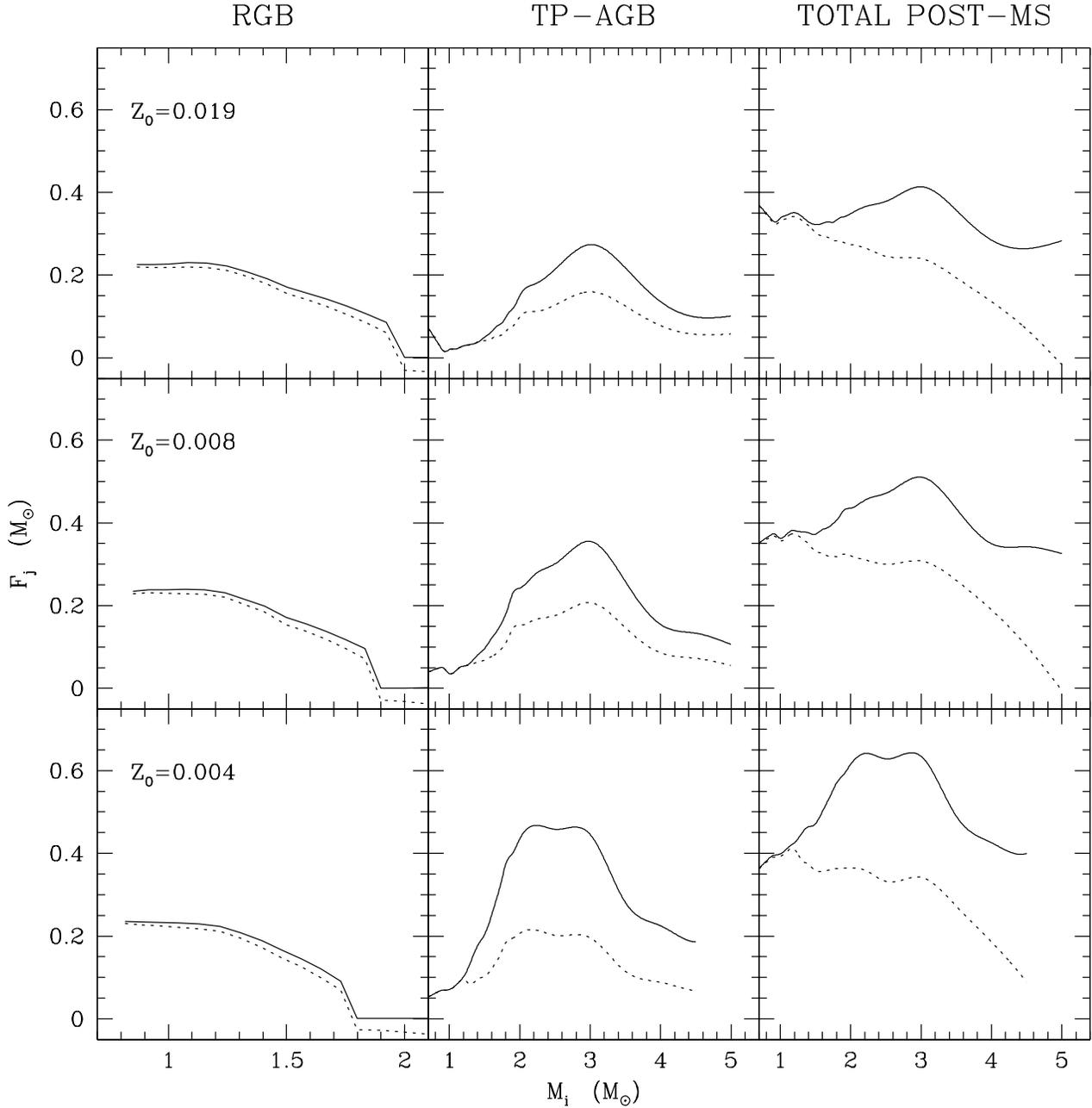}}
\caption{Stellar fuel consumption (derived with the analytical
formalism) of low- and intermediate-mass models with different choices
of the initial metallicity, and referring to i) the RGB phase (left
panels); TP-AGB phase (middle panels); and total post-MS evolution
(right panels).  Results of calculations including the effect of
dredge-up events and hot-bottom burning (solid lines) are compared to
the case in which such effect is neglected (dotted lines).  See text
for more details}
\label{fig_2}
\end{figure*}
     
As illustrated in Fig.~\ref{fig_1} the differences between the two
cases are remarkable for this particular set of stellar models. As
expected, the total fuel consumption according to case A is
systematically underestimated with respect to case B.  In particular,
the following points are worthy of notice.
	\begin{itemize} 
	\item The discrepancy -- between the results obtained in case
A and the correct ones with prescription B -- increases, on average,
with the stellar mass, reflecting the trend of the chemical yields
(bottom panel).
	\item Lower mass stars (with $M \sim 1 \, M_{\odot}$)
experience only the first dredge-up during their evolution, so that
the ``chemical'' terms almost do not contribute to their $F_{\rm T}$.
	\item At around $M \sim 3 M_{\odot}$ both the actual total
fuel $F_{\rm T}$ -- calculated according to case B --, and the yield
of helium, $M_{\rm y}$(He), present a maximum, which corresponds to
the longest duration of the TP-AGB phase for this set of stellar
models (see Marigo 2001).
	\item At larger stellar masses, prescription A predicts a
drastic drop of the fuel, which does not show up with prescription
B. This can be explained comparing the curves of $\Mcore^{\rm TO}$ and
$\Mcore^{\rm AGBt}$ (middle panel of Fig.~\ref{fig_1}). The difference
$(\Mcore^{\rm AGBt} - \Mcore^{\rm TO})$ progressively reduces at
increasing stellar mass, eventually becoming negative.  In other
words, in the most massive models the final cores left at the end of
the AGB phase may have even lower masses that the cores built up at
the end of the MS.  In these cases, the fuel calculated with
prescription A would eventually attain negative values, thus losing
significance.  This example is already a clear indication that the
evaluation of the stellar fuel simply basing on core masses is not a
correct procedure, as it may even produce meaningless results.
Conversely, looking at predictions obtained with prescription B, we
can see the total fuel first decreases and then flattens towards the
most massive models (keeping always positive), due to the combined
effect of both $M_{\rm c}$ and chemical yields.
	\end{itemize} 

The weight of the ``chemical terms'' can be appreciated also in
Fig.~\ref{fig_2}, which displays the results of the analytical
estimation of the fuel consumed during the RGB, TP-AGB and total
post-MS phases, as a function of the initial stellar mass and
metallicity.  We can see that the effect of the $1^{\rm st}$ dredge-up
occurring on the RGB is rather small, whereas accounting or not for
the surface chemical changes during the AGB leads to quite different
results, i.e. the fuel can be substantially underestimated if not
doing correctly.  Moreover, we notice that the stellar fuel
consumption presents a marked {\sl metallicity dependence}, i.e.  it
increases, on average, at decreasing metallicity for all the
evolutionary phases here considered.

It is also clear that the stellar fuel -- calculated with the proposed
formalism -- is positively correlated to the efficiency of the
dredge-up episodes.  To give an example, let us consider the TP-AGB
evolution of a given stellar model. For the sake of simplicity, we
assume that at each thermal pulse a dredge-up occurs with a constant
efficiency, $\lambda = \Delta M_{\rm dred}/\Delta M_{\rm c}$, defined
as the fraction of the core mass increment during an inter-pulse
period which is dredged-up into the envelope at the subsequent thermal
pulse.  This implies that, every time a dredge-up takes place, the
core mass is reduced by the amount $\lambda \Delta M_{\rm c}$.  The
TP-AGB fuel consumption can be consequently expressed as:
\beq
F_{\rm TP-AGB}  \simeq (X_{1,2} + 0.1)\, 
\frac{(\Mcore^{\rm AGBt} - \Mcore^{\rm 1TP})}{(1-\lambda)} 
\eeq
that is equivalent to Eq.~(\ref{eq_agbfuel}).

It follows that, if we neglect the effect of third dredge-up, the
stellar TP-AGB fuel consumption would be underestimated by a factor
$1/(1-\lambda)$. Setting $\lambda = 0.50$, as adopted in the
calculations presented in Fig.~\ref{fig_1}, the underestimation factor
is $\sim 2$, i.e. the actual TP-AGB fuel would be twice larger than
that one would obtain setting $\lambda=0$.  Finally, it is worth
noticing that such effect may be even larger if one assumes the
extreme dredge-up efficiencies ($\lambda \ga 1$) of recent TP-AGB
calculations (i.e. Herwig 2000).

Of course, only case B represents the correct approach to be followed
in order to estimate the stellar fuel.  However, discussing case A is
still useful.  In fact, case A represents a sort of {\em lower limit
to the fuel}, that depends only on the core masses at the MS and AGB
termination stages. Since $\Mcore^{\rm TO}$ is a monotonic function of
the initial mass $M_{\rm i}$, and $\Mcore^{\rm AGBt}=M_{\rm WD}$, it
follows that this lower limit is intimately linked to the $M_{\rm
WD}(M_{\rm i})$ relation (see also Girardi \& Bertelli 1998). This
latter is known as the initial--final mass relation (IFMR), and is
well constrained by observations of nearby white dwarfs (both in the
fields and in open clusters). Thus, the observed IFMR poses a direct
constraint -- i.e. a lower limit -- to the behaviour of the stellar
fuel as a function of the mass.

In the context of galaxy models, case A can be seen as {\em the fuel}
(intended as the luminosity integral of Eq.~\ref{eq_flum}) {\em that
we should be dealing with in our spectro-photometric model} -- instead
of the one given by case B -- {\em under the assumption that low- and
intermediate-mass stars do not contribute to the chemical evolution of
the system under consideration}.  Needless to say, this situation is
not realistic at all.

\subsection{The need for a consistency choice in galaxy models}
\label{sec_theneed}

The analysis and relative discussion carried out in the previous
sections should have already convinced the reader on the intimate
connection between light emission and chemical yields from stars.
Up to now, we have illustrated this fact using a homogeneous set
of stellar models (Girardi et al. 2000; Marigo et al. 1999, and Marigo
2001).  To make concepts even clearer, let us suppose to opt for an
{\em inconsistent} choice in modelling a galaxy, that is combining the
TP-AGB tracks of Marigo (2001, M2K; see~Appendix A) to model the
spectrophotometric evolution, with the stellar yields of Renzini \&
Voli (1981, RV81) to model the chemical evolution\footnote{
Together with those of van den Hoek \& Groenewegen (1997), M2K and
RV81 yields are commonly used in chemical evolution models of
galaxies. A detailed comparison among the different set of stellar
yields is provided in Marigo (2001), to whom we refer for all
details.}

In this case, we can get two different estimates of the fuel: the
first one is derived from our analytical formulas applied to RV81
yields and core masses and corresponds to fuel adopted in the chemical
evolution model, and the second one comes from the exact method
applied to M2K evolutionary tracks and corresponds to fuel adopted in
the spectrophotometric model. Our primary aim is to see how they
compare. If the two different inputs were consistent, the two
estimates of the fuel should agree to within a few percent, just as in
the case of Fig.~\ref{fig_3}. Otherwise, this exercise would allow us
to quantify the degree of mismatch (inconsistency) between these
two sources of stellar data.

\begin{figure*}
\resizebox{\hsize}{!}{\includegraphics{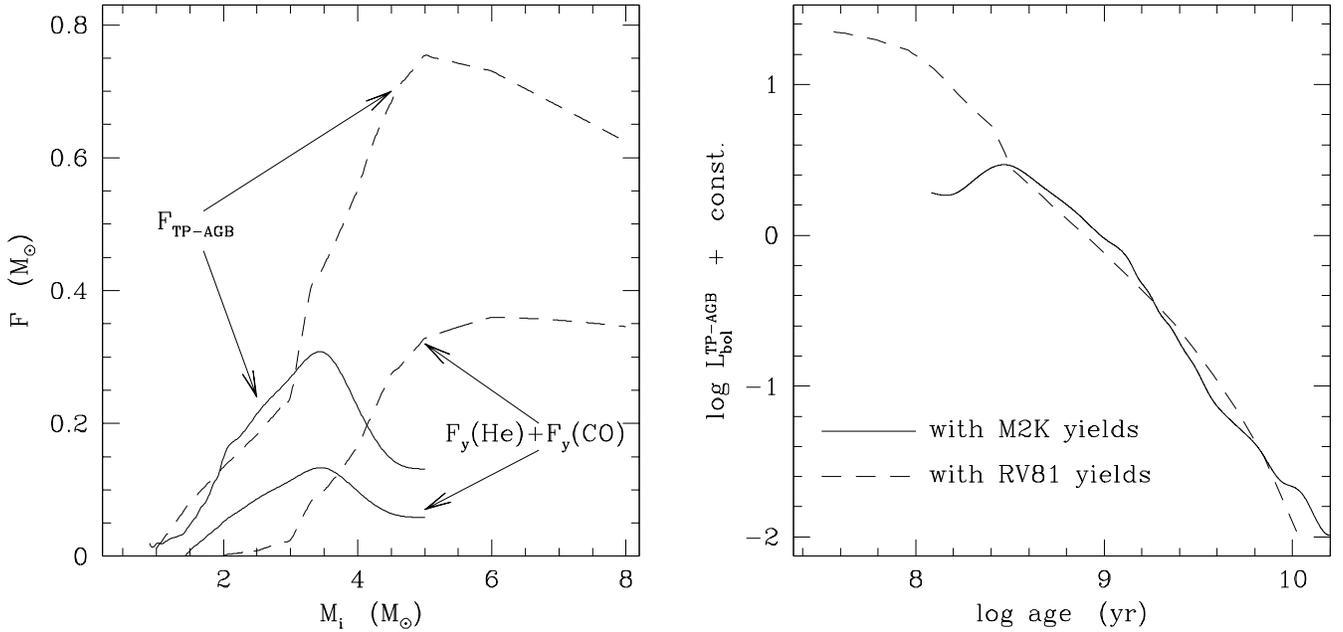}}
\caption{Left panel: TP-AGB fuel calculated with the analytical
formalism here presented and applied to stellar models of Renzini \&
Voli (1981, RV81; dashed line), and Marigo (2001, M2K; continuous line).  
Right panel: Contribution (in arbitrary units) of the TP-AGB phase to the 
integrated bolometric luminosity of SSPs as a function of age. 
For RV81 models the evolutionary flux $b(t)$ is calculated following 
Maraston (1998), whereas for M2K models we base on Girardi et al. (2000) 
evolutionary calculations, in both cases with the assumption of the Salpeter 
IMF. See text for more details.}
\label{fig_4}
\end{figure*}

We use Equation~(\ref{eq_agbfuel}) to derive the fuel consumption of
TP-AGB models with initial metallicity $Z_0=0.02$ presented by RV81.
The results are shown in Fig.~\ref{fig_4} (left panel). We adopt the
final core masses ($\Mcore^{\rm AGBt}$) and chemical yields presented
in tables 3f and 3e of RV81, this latter table corresponding to
calculations of TP-AGB models (with $M \ge 3.3 \, M_{\odot}$) with
hot-bottom burning carried out with the mixing-length parameter
$\alpha=2.0$.  For RV81 models, the amounts of mass lost during the
RGB phase of low-mass stars (to be properly subtracted from the total
yields to get the TP-AGB yields), the envelope abundances of $^{4}$He
(necessary to calculate the quantities $Y'$), and the corresponding
core masses at the onset of the TP-AGB phase ($\Mcore^{\rm 1TP}$) are
derived following the recipes described in their section 2.

For M2K models with $Z_0=0.019$ and $\alpha=1.68$, the TP-AGB
fuel has been already presented in our previous
Sects.~\ref{sec_accuracy} (see Fig.~\ref{fig_3}).  The left panel of
Fig.~\ref{fig_4} shows, instead, M2K fuels corresponding to the same
value of $\alpha=2.0$ as adopted in the RV81 models presented in the
same figure.

For both sets of stellar models, we also compute the 
TP-AGB integrated bolometric luminosity as a function of SSP age,
according to:
	\beq
L_{\rm bol}^{\rm TP-AGB} = A_{\rm H}\, b(t)\, F_{\rm TP-AGB}(M_{\rm TO}) 
 	\eeq 
(where all quantities are defined in Sect.~\ref{sec_ssp}). They
are displayed in the right panel of Fig.~\ref{fig_4}.
 
It is clear, from the comparison between RV81 and M2K results, that
the TP-AGB fuel of RV81 models is much larger than the one of M2K
models, for masses $M_{\rm i}\ga 3 \, M_\odot$, corresponding to
stellar ages shorter than about 300~Myr. More modest but still
non-negligible differences are also present at lower masses (older
ages).

In addition to the TP-AGB fuel as a function of the initial stellar
mass, Fig.~\ref{fig_4} (left panel) shows also the contribution of the
``chemical'' terms to $F_{\rm TP-AGB}$ (i.e. the second and third
terms in Eq.~\ref{eq_agbfuel}) for both sets of models.  It turns out
that i) in both cases the yield contribution to the fuel is generally
relevant (except for the lowest mass models that do not experience
both the third dredge-up and hot-bottom burning), ii) the differences
between RV81 and M2K are still considerable.  We can notice that for
$M_{\rm i} \ga 4 \, M_{\odot}$ the yield contribution to the TP-AGB
fuel in RV81 is even larger than the total TP-AGB fuel in M2K models.

As expected, the trend of $F_{\rm TP-AGB}$ as a function of $M_{\rm
i}$ (left panel of Fig.~\ref{fig_4}) reflects in the behaviour of
$L_{\rm bol}^{\rm TP-AGB}$ as a function of the corresponding SSP age
(for $M_{\rm TO} = M_{\rm i}$; right panel of Fig.~\ref{fig_4}).  In
fact, here the most relevant differences between RV81 and M2K
predictions show up at the youngest ages (hence higher $M_{\rm TO}$),
i.e.  the luminosity contribution of the TP-AGB phase according to
RV81 models is much greater than predicted by M2K models.
 
On the basis of the above comparative analysis, let us now
consider what happens if M2K tracks are used to model the
spectro-photometric evolution, whereas RV81 yields are used to model
the chemical evolution. The most striking inconsistency in this model
would show up at $M_{\rm i} \sim 4\, M_{\odot}$  (or equivalently
at $\log t\la 8.4$), where the fuel necessary to account for RV81
yields would already be larger than the fuel necessary to account for
the luminosity of M2K models.  For TP-AGB stars in such a range of
masses and ages, our galaxy model we would be either strongly
overestimating the chemical yields or, equivalently, significantly
underestimating the emitted light.  The degree of inconsistency is
smaller, but still significant, at smaller masses (longer ages).  

Since the TP-AGB phase provides a significant fraction of the total
energy emitted by stellar populations (meant either as bolometric
light, or as chemical yields), one may deduce {\em that such models
can hardly be considered as self-consistent in terms of their
energetics}.

Moreover, a number of other subtle inconsistencies may affect such a
galaxy model. For instance, RV81 and M2K models predict quite
different initial--final mass relations (see M2K for a
discussion). Whereas a chemical model that uses RV81 yields is
expected to produce a relatively large fraction of ``massive'' white
dwarfs (with $M_{\rm WD}\sim 1\,M_\odot$), a spectro-photometric model
that adopts M2K tracks, under the same assumptions for star formation
rate and age-metallicity relation, is expected to predict much fewer
of them.  Then, in view of modelling, for instance, the formation of
galactic halos one would face the embarrassing question: which one of
the two theoretical white dwarf mass distributions should be
considered for a comparison with the observational data of field WDs
and MACHOs\,? Any is the answer, it will be the wrong one.

\section{Final remarks}

This study has highlighted the intimate relation between the
stellar emitted light and chemical yields.  It has been clarified that
a significant fraction of fuel consumption (emitted light) may be
eventually ``deposited'' in the form of He and CO yields. This
fraction can be quantified by means of simple analytical relations.
We have illustrated this in detail for low- and intermediate-mass
stars, also suggesting how the same analysis could be extended to
massive stars.  Then, the derived formalism offers a useful
consistency check that can be applied to galaxy models whenever
different sources of stellar data are employed to model the chemical
and spectrophotometric evolution.

Such consistency check expresses essentially the basic condition of
energy conservation, which can be read as: the emitted light implies a
precise amount of stellar nucleosynthesis, part of which corresponds
to chemical yields.  Other fundamental aspects, such as mass
conservation, are already included in the basic equations of chemical
evolution models.
 
It is important to remark that the present analysis is general, and
independent of the specific set of stellar tracks/yields in use.  As a
matter of fact, analysing the quality of any given set of stellar
(or galaxy) models, is beyond the scope of this paper.

Anyway, a few remarks can be made. As indicated by our
analysis, the primary yields of He and CNO elements are those more
directly related to the emitted light of a stellar population.  In
galactic chemical evolution models, He and CNO yields critically
determine the evolution of $Y(t)$ (or, equivalently, the $\Delta
Y/\Delta Z$ ratio) and $Z(t)$ relations.  Our formalism suggests
that, if a given set of stellar models predicts a particular evolution
of the integrated light, {\em the same set of models} necessarily
implies a particular evolution of $Y(t)$ and $Z(t)$. 

Of course, the situation is somewhat complicated by the fact that most
of the metals are produced by massive stars, which are characterised
by rapid evolution, so that soon they do not contribute to the
integrated light anymore. Hence, light and chemical evolution may be
dominated by different stars, possibly justifying the approximation
that photometric and chemical evolution could be treated separately.
Actually, this assumption may be safely applied to some cases (e.g.\
in the study of the present Galaxy), but becomes questionable (and
risky) once {\em galaxy models are constructed to describe the
evolution of galaxies over most of their history, since their
formation up to present times}. An example of this ``unifying''
approach is given by the studies of high-redshift galaxies based on
models ``calibrated'' on local galaxy samples.  Clearly, our analysis
indicates that the energetics necessary to explain the light of the
distant sample has strict implications for the chemistry now observed
in the local sample, and vice-versa.

Finally, we would like to conclude outlining the main points of the
present work:
	\begin{itemize}
	\item The stellar fuel (emitted light) consumed by a star
during a specified evolutionary phase may be schematically decomposed
as the sum of two terms: a contribution {\sl locked} in the star and
related to the mass of the core regions and envelope composition, and
a contribution {\sl emitted} by the star during that phase in the form
of chemical yields.  These terms are explicited in this paper by means
of suitable and simple analytical formulas, covering various phases of
stellar evolution.
	\item When coupling the spectro-photometric and chemical
evolution of a system, we should care that the basic contributions of
the stellar component -- i.e. light and newly synthesised elements --
are included in the models consistently one with each other,
preferably derived from the same set of stellar models.  If this is
not the case, i.e. the input stellar data come from heterogeneous
sources, we could at least measure the {\sl degree of internal
mismatch} of the model. The analytical formalism developed in this
work can serve the purpose.
	\end{itemize}


\subsection*{ Acknowledgements }

L.G.\ thanks Martin Groenewegen for very useful conversations about
the third dredge-up.  We warmly thank Elena Fantino for her invaluable
help in the preparation of the figures.  This work was partly funded
by the Italian Ministry of University, Scientific Research and
Technology (MURST).

\section*{ References }

\begin{description}
\item Arimoto N., Yoshii Y., 1986, A\&A 164, 260
\item Arimoto N., Yoshii Y., 1987, A\&A 173, 23
\item Bertelli G., Bressan A., Chiosi C., Fagotto F, Nasi E., 
      1994, A\&AS 106, 275
\item Bressan A., Chiosi C., Fagotto F., 1994, ApJS 94, 63
\item Chabrier G., 1999, ApJ 513, L103
\item Chandrasekhar. S., 1939,  
	An Introduction to the Study of Stellar Structure,
	The University of Chicago Press, Chicago
\item Charlot S., Silk J., 1995, ApJ 445, 124
\item Chiosi C., Bressan A., Portinari L., Tantalo R., 1998, A\&A 339, 355
\item Einsel C., Fritze-von Alvensleben U., Krueger H., Fricke K.J., 1995,
	A\&A 296, 347
\item Fall S.M., 2001, in The Extragalactic Infrared Background and its
     Cosmological Implications, IAU Symp. 204, 
     Eds. M. Harwit and M.G. Hauser, in press (astro-ph/0101084). 
\item Ferreras I., Silk J., 2000, MNRAS 316, 786
\item Fields B.D., Freese K., Graff D.S., 2000, ApJ 534, 265
\item Frogel J.A., Mould J., Blanco V.M., 1990, ApJ 352, 96
\item Girardi L., Bertelli G., 1998, MNRAS 300, 533
\item Girardi L., Bressan A., Bertelli G., Chiosi C., 2000, A\&AS  141, 371
\item Greggio L., Renzini A., 1990, ApJ 364, 35
\item Herwig F., 2000, A\&A 360, 952
\item Ibata R., Irwin M., Bienaym\'e O., Scholz R., Guibert J, 2000,
	ApJ 532, L41
\item Madau P., Pozzetti L., 2000, MNRAS 312, L9  
\item Maeder A., 1992, A\&A 264, 105
\item Maraston, C., 1998, MNRAS 300, 872 
\item Marigo P., Girardi L., Bressan A., 1999, A\&A 344, 123
\item Marigo P., 2001, A\&A 370, 194
\item Portinari L., Chiosi C., Bressan A., 1998, A\&A 334, 505
\item Renzini A., Buzzoni A., 1986, in Spectral Evolution of Galaxies, 
        eds.\ C.\ Chiosi and A.\ Renzini, Dordrecht: Reidel, p.\ 195
\item Renzini A., Voli M., 1981, A\&A 94, 175
\item Renzini A., 1998, AJ 115, 2459
\item Tinsley B.M., 1980, Fund. Cosmic Phys. 5, 287
\item Tuchman Y., Glasner A., Barkat Z., 1983, ApJ 268, 356
\item van den Hoek L.B., Groenewegen M.A.T., 1997, A\&AS 123, 305
\item Vazdekis A., Casuso E., Peletier R.F., Beckman J.E., 1996,
	ApJS, 106, 307
\item Woosley S.E., Weaver T.A., 1995, ApJS 101, 181  
\end{description}

\appendix
\section{Isochrones with improved TP-AGB models}
\label{sec_isoc}

Girardi et al. (2000) presented a large set of evolutionary tracks for
low- and intermediate-mass stars, covering the mass range from $0.15$
to $7\,M_\odot$, for six different initial chemical
compositions. These models account for a moderate amount of convective
overshooting from stellar cores, which makes the upper mass limit for
the development of the AGB phase to be about $5\,M_\odot$, instead of
$8\,M_\odot$ as predicted by classical stellar models.

Since the evolutionary calculations, carried out by means of the
complete stellar code, stopped after the first few thermal pulses,
Girardi et al. (2000) followed the subsequent TP-AGB phase with the
aid of a simple synthetic algorithm (cf.\ Girardi \& Bertelli
1998). In this way, sets of isochrones including the complete TP-AGB
phase were presented.

That approach to the TP-AGB evolution, however, is far too simple if
compared to the detailed semi-analytical TP-AGB models presented by
Marigo et al. (1999, and references therein).  These latter models
include the most crucial aspects of TP-AGB evolution, such as the
third dredge-up, hot-bottom burning, deviations from the core
mass--luminosity relation, and mass-loss rates related to the
pulsational period.  Moreover, the basic model parameters have been
calibrated such as to reproduce the carbon star luminosity functions
in the LMC and SMC. Chemical yields from these models are presented by
Marigo (2001).

Marigo et al. (1999) tracks start at the first thermal pulse computed
by Girardi et al.\ (2000), so that both sets of models are perfectly
complementary.  They follow the entire TP-AGB evolution up to the
complete ejection of the stellar envelope. These synthetic TP-AGB
tracks are available, so far, only for chemical compositions
$[Z_0=0.004, Y_0=0.240]$, $[Z_0=0.008, Y_0=0.250]$ and $[Z_0=0.019,
Y_0=0.273]$. For these metallicities, the synthetic TP-AGB models have
been suitably combined to the tracks calculated by Girardi et al.\
(2000), so that theoretical isochrones have been constructed using the
same method as in Girardi et al.\ (2000).

Thanks to their continuity and homogeneity, the present isochrones are
well suited for the population synthesis of galaxies.  Moreover, they
perfectly meet the consistency requirements discussed in this paper:
the emitted light computed with the present isochrones (with the aid
of Eq.~\ref{eq_isochrone}), would be completely consistent with the
chemical evolution derived from the tables of chemical yields and
remnant masses in Marigo (2001).

\begin{figure}
\resizebox{\hsize}{!}{\includegraphics{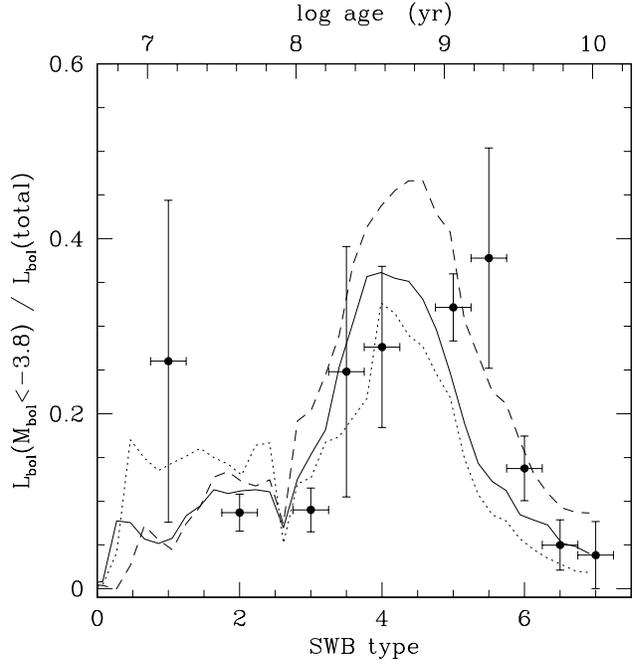}}
\caption{The fractional contribution of luminous AGB stars tho the
integrated light of LMC and SMC star clusters, as inferred from Frogel
et al. (1990) data (full dots with error bars), as a function of the
equivalent SWB types. This latter represents a rough age ranking for
LMC and SMC clusters, that corresponds to the logarithmic age scale
presented in the upper axe.  The same quantity is plotted, as a
function of age, as derived from our isochrones of metallicity
$Z_0=0.004$ (dashed line), $Z_0=0.008$ (continuous line) and
$Z_0=0.019$ (dotted line). }
\label{fig_swb}
\end{figure}

Moreover, it is worth emphasizing that Marigo et al.\ (1999) models
reproduce a series of observational properties of AGB stars in nearby
galaxies, and of nearby white dwarfs (see Marigo 2001 for details).
Therefore, one may expect that they are adequate to model the
integrated light emitted by TP-AGB stars.

\subsection{Integrated ligth contribution from AGB stars}

In order to check this latter point, we compare the present models to
the data from Frogel et al.\ (1990), who measured the contribution of
luminous AGB stars to the integrated light of Magellanic Cloud star
clusters of various ages. To make the comparison, we proceed as
follow: First, from their table 1 we compute the bolometric light
coming from stars with $M_{\rm bol}<-3.8$ in each cluster (assuming
distance moduli of 18.5 and 18.9~mag for the LMC and SMC,
respectively).  This $M_{\rm bol}$ limit guarantees that we are
dealing only with AGB stars brighter than the RGB-tip.  Second, we
divide the clusters into bins of equivalent SWB (Searle et al. 1980)
types (cf. table 4 in Frogel et al.).  Third, we compute the ratio,
$L_{\rm bol}(M_{\rm bol}<-3.8) / L_{\rm bol}({\rm total})$, between
the luminosity coming from stars with $M_{\rm bol}<-3.8$ and that
emitted from the whole cluster (cf. their table 4).  Finally, we
estimate the corresponding error bars from the total number of $M_{\rm
bol}<-3.8$ stars in each SWB bin.

The results are presented in Fig.~\ref{fig_swb}, which also displays
the approximate relation between SWB type and the logarithm of cluster
age, i.e. $\log(t/{\rm yr})\simeq 6.66+0.48\,{\rm SWB}$.  The observed
data in this figure are indeed very similar to the original ones in
figure 15 of Frogel et al. (1990).  Superimposed on the data, we plot
our theoretical expectation for the same quantity, $L_{\rm bol}(M_{\rm
bol}<-3.8) / L_{\rm bol}({\rm total})$, as derived from the present
isochrones.

From the figure, it is well evident that the models correctly
reproduce the general behaviour of the relative luminosity
contribution over the complete age interval in which AGB stars are
present (i.e. $8 \la \log(t/{\rm yr}) \la 10$).  In particular, the
minimum values at $\log(t/{\rm yr}) \sim 8$ and $\sim 10$ are clearly
present in the models, as well as the maximum at $\log(t/{\rm yr})
\sim 8.6$, that would correspond to the longest TP-AGB lifetimes (see
Marigo 2001).  It is also worth noticing that according to the models
at decreasing metallicity the luminosity contribution from AGB stars
increases.

For ages younger than $10^8$~yr, we have computed the same quantities
using Bertelli et al. (1994) isochrones. The integration of $L_{\rm
bol}(M_{\rm bol}<-3.8)$ was limited to stars with $\log T_{\rm
eff}<3.8$, in order to exclude the blue main sequence stars from this
term.  As shown in Fig.~\ref{fig_swb}, the models predict values
$L_{\rm bol}(M_{\rm bol}<-3.8) / L_{\rm bol}({\rm total})$ of around
0.1 at these young ages. They are in reasonable agreement with
observations, despite of the fact that the empirical data present a
slight hint for an increase in $L_{\rm bol}(M_{\rm bol}<-3.8)$ at even
younger ages ($\sim 10^7$~yr) (we notice, however, this feature is
based on too few stars to be completely reliable).

With the cautionary remarks that (i) the SWB classification represents
just a crude age ranking for clusters, whose consistency with our
models has not been checked for, and (ii) that there is plenty of room
for improving the empirical data for AGB stars in clusters, we
conclude that our models reproduce the observed contribution of AGB
stars to the integrated light of SSPs in a satisfactory way.


\subsection{Useful data}

\input{fuelz02_al168.tex}
\input{fuelz008_al168.tex}
\input{fuelz004_al168.tex}

Complete tables with the improved isochrones described in the previous
section are available upon request to the authors, and at the WWW site
\verb$http://pleiadi.pd.astro.it$.

For the present models, Tables A1 -- A3 contain the stellar fuels as a
function of the initial mass, for various post-MS phases. All fuels
have been calculated by means of Eq.~(\ref{eq_flum}).

We also present fitting relations to $\Mcore^{\rm TO}$ as a function
of the initial mass $M_{\rm i}$ and metallicity $Z_0$, derived from
the stellar models here considered (with overshooting) :
	\beqa
	\Mcore^{\rm TO} & \simeq & 0.463 - 0.654\, M_{\rm i} + 0.338\,  M^2_{\rm i} \\
\nonumber
               & & -0.024 \, \log(Z_0/0.019)\,\,\,\,\,\,\,\,\,\,\,\,\,\,\,\,\,\,\,\,\,\,
	{\rm for}\, M_{\rm i} \la 1.4 \\
	\Mcore^{\rm TO} & \simeq & 0.006 + 0.115 \, M_{\rm i} + 0.023\,  M^2_{\rm i} \\
\nonumber
               & & -0.057 \, \log(Z_0/0.019)\,\,\,\,\,\,\,\,\,\,\,\,\,\,\,\,\,\,\,\,\,\,
	{\rm for}\, M_{\rm i} \ga 1.4 \, ,
	\eeqa
and from classical models (without overshooting) with initial solar composition 
($Y_0=0.273\, , Z_0=0.019$): 

	\beqa
	\Mcore^{\rm TO} & \simeq & 0.123 + 0.066 \, M_{\rm i} - 0.037\,  M^2_{\rm i} \\
\nonumber 
	& & \,\,\,\,\,\,\,\,\,\,\,\,\,\,\,\,\,\,\,\,\,\,\,\,\,\,\,\,\,\,\,\,\,
	\,\,\,\,\,\,\,\,\,\,\,\,\,\,\,\,\,\,\,\,\,\,\,\,\,\,\,\,\,\,\,\,\,\,\,
	\,\,\,\,\,\,\,\,\,\,\,\,
	{\rm for}\, M_{\rm i} \la 1.4\\
	\Mcore^{\rm TO} & \simeq & 0.015 + 0.063 \, M_{\rm i} + 0.019\,  M^2_{\rm i}\\
\nonumber
         & & \,\,\,\,\,\,\,\,\,\,\,\,\,\,\,\,\,\,\,\,\,\,\,\,\,\,\,\,\,\,\,\,
	\,\,\,\,\,\,\,\,\,\,\,\,\,\,\,\,\,\,\,\,\,\,\,\,\,\,\,\,\,\,\,\,\,\,\,
	\,\,\,\,\,\,\,\,\,\,\,\,\,
	{\rm for}\, M_{\rm i} \ga 1.4 \, .
	\eeqa
All masses are expressed in solar units.  We remind that according to
our definition of $\Mcore^{\rm TO}$ (see Eq.~\protect\ref{eq_mcto}),
the fuel burnt during the MS can be estimated from:
	\beq
	F_{\rm MS} = X_0\, \Mcore^{\rm TO}
	\eeq

\end{document}

%% file: fuelz02_al168.tex
\begin{table}
\begin{tabular}{ccccc}
\multicolumn{5}{l}{{{\bf Table A1.} Post-MS stellar fuels
($M_{\odot}$) -- $Z_0=0.019$}} \\
\noalign{\smallskip}
\hline
\noalign{\smallskip}
\multicolumn{1}{c}{$M_{\rm i}$} &
\multicolumn{1}{c}{$F_{\rm RGB}$} &
\multicolumn{1}{c}{$F_{\rm CHeb + E-AGB}$} &
\multicolumn{1}{c}{$F_{\rm TP-AGB}$} &
\multicolumn{1}{c}{$F_{\rm T}$} \\
\noalign{\smallskip}
\hline
\noalign{\smallskip}
0.868  &   2.371E-01  &   1.001E-01  &  2.892E-02  &   3.661E-01\\
0.934  &   2.387E-01  &   1.014E-01  &  1.400E-02  &   3.542E-01\\
1.005  &   2.405E-01  &   1.023E-01  &  1.912E-02  &   3.620E-01\\
1.082  &   2.428E-01  &   1.059E-01  &  1.887E-02  &   3.675E-01\\
1.163  &   2.385E-01  &   1.042E-01  &  2.402E-02  &   3.667E-01\\
1.248  &   2.287E-01  &   1.061E-01  &  2.778E-02  &   3.625E-01\\
1.334  &   2.141E-01  &   1.091E-01  &  3.049E-02  &   3.537E-01\\
1.420  &   1.951E-01  &   1.095E-01  &  3.577E-02  &   3.405E-01\\
1.504  &   1.731E-01  &   1.158E-01  &  4.829E-02  &   3.372E-01\\
1.588  &   1.573E-01  &   1.182E-01  &  5.916E-02  &   3.347E-01\\
1.672  &   1.415E-01  &   1.179E-01  &  7.404E-02  &   3.334E-01\\
1.756  &   1.230E-01  &   1.245E-01  &  8.503E-02  &   3.326E-01\\
1.839  &   1.006E-01  &   1.327E-01  &  1.065E-01  &   3.398E-01\\
1.923  &   6.746E-02  &   1.483E-01  &  1.222E-01  &   3.380E-01\\
2.000  &   1.910E-02  &   1.870E-01  &  1.485E-01  &   3.546E-01\\
2.200  &   1.640E-02  &   1.781E-01  &  1.766E-01  &   3.711E-01\\
2.500  &   1.130E-02  &   1.652E-01  &  2.145E-01  &   3.909E-01\\
3.000  &   8.800E-03  &   1.597E-01  &  2.693E-01  &   4.378E-01\\
3.500  &   8.600E-03  &   1.604E-01  &  2.149E-01  &   3.839E-01\\
4.000  &   9.700E-03  &   1.780E-01  &  1.363E-01  &   3.239E-01\\
4.500  &   1.090E-02  &   1.944E-01  &  1.001E-01  &   3.052E-01\\
5.000  &   1.170E-02  &   2.144E-01  &  1.030E-01  &   3.289E-01\\
\noalign{\smallskip}
\hline
\end{tabular}
\end{table}

%% file: fuelz008_al168.tex
\begin{table}
\begin{tabular}{ccccc}
\multicolumn{5}{l}{{{\bf Table A2.} Post-MS stellar fuels
($M_{\odot}$) -- $Z_0=0.008$}} \\
\noalign{\smallskip}
\hline
\noalign{\smallskip}
\multicolumn{1}{c}{$M_{\rm i}$} &
\multicolumn{1}{c}{$F_{\rm RGB}$} &
\multicolumn{1}{c}{$F_{\rm CHeb + E-AGB}$} &
\multicolumn{1}{c}{$F_{\rm TP-AGB}$} &
\multicolumn{1}{c}{$F_{\rm T}$} \\
\noalign{\smallskip}
\hline
\noalign{\smallskip}
0.850  &   2.542E-01  &   9.550E-02  &  4.851E-02  &   3.982E-01\\
0.918  &   2.522E-01  &   9.746E-02  &  4.888E-02  &   3.986E-01\\
0.992  &   2.516E-01  &   1.006E-01  &  3.666E-02  &   3.889E-01\\
1.071  &   2.514E-01  &   1.017E-01  &  4.052E-02  &   3.936E-01\\
1.154  &   2.499E-01  &   1.048E-01  &  5.144E-02  &   4.061E-01\\
1.239  &   2.414E-01  &   1.066E-01  &  5.500E-02  &   4.030E-01\\
1.407  &   2.061E-01  &   1.055E-01  &  7.898E-02  &   3.906E-01\\
1.499  &   1.800E-01  &   1.101E-01  &  9.679E-02  &   3.867E-01\\
1.583  &   1.626E-01  &   1.138E-01  &  1.172E-01  &   3.936E-01\\
1.667  &   1.452E-01  &   1.168E-01  &  1.392E-01  &   4.012E-01\\
1.750  &   1.235E-01  &   1.224E-01  &  1.644E-01  &   4.102E-01\\
1.832  &   9.532E-02  &   1.347E-01  &  1.985E-01  &   4.285E-01\\
1.900  &   2.450E-02  &   1.866E-01  &  2.317E-01  &   4.427E-01\\
2.000  &   2.380E-02  &   1.850E-01  &  2.451E-01  &   4.540E-01\\
2.200  &   1.720E-02  &   1.808E-01  &  2.758E-01  &   4.738E-01\\
2.500  &   1.260E-02  &   1.759E-01  &  3.047E-01  &   4.933E-01\\
3.000  &   1.020E-02  &   1.871E-01  &  3.539E-01  &   5.513E-01\\
3.500  &   9.700E-03  &   2.116E-01  &  2.581E-01  &   4.794E-01\\
4.000  &   1.030E-02  &   2.288E-01  &  1.552E-01  &   3.943E-01\\
4.500  &   1.080E-02  &   2.426E-01  &  1.354E-01  &   3.888E-01\\
5.000  &   1.150E-02  &   2.589E-01  &  1.083E-01  &   3.787E-01\\
\noalign{\smallskip}
\hline
\end{tabular}
\end{table}

%% file: fuelz004_al168.tex
\begin{table}
\begin{tabular}{ccccc}
\multicolumn{5}{l}{{{\bf Table A3.} Post-MS stellar fuels 
($M_{\odot}$) -- $Z_0=0.004$}} \\
\noalign{\smallskip}
\hline
\noalign{\smallskip}
\multicolumn{1}{c}{$M_{\rm i}$} &
\multicolumn{1}{c}{$F_{\rm RGB}$} &
\multicolumn{1}{c}{$F_{\rm CHeb + E-AGB}$} &
\multicolumn{1}{c}{$F_{\rm TP-AGB}$} &
\multicolumn{1}{c}{$F_{\rm T}$} \\
\noalign{\smallskip}
\hline
\noalign{\smallskip}
0.817  &   2.536E-01  &   9.664E-02  &  5.876E-02  &   4.089E-01\\
0.888  &   2.527E-01  &   1.005E-01  &  6.502E-02  &   4.182E-01\\
0.966  &   2.517E-01  &   1.040E-01  &  6.470E-02  &   4.204E-01\\
1.046  &   2.499E-01  &   1.067E-01  &  7.085E-02  &   4.274E-01\\
1.131  &   2.467E-01  &   1.091E-01  &  8.151E-02  &   4.373E-01\\
1.218  &   2.397E-01  &   1.115E-01  &  9.972E-02  &   4.510E-01\\
1.307  &   2.223E-01  &   1.122E-01  &  1.392E-01  &   4.736E-01\\
1.396  &   2.002E-01  &   1.147E-01  &  1.771E-01  &   4.919E-01\\
1.481  &   1.741E-01  &   1.153E-01  &  1.989E-01  &   4.882E-01\\
1.564  &   1.508E-01  &   1.198E-01  &  2.360E-01  &   5.066E-01\\
1.647  &   1.246E-01  &   1.275E-01  &  2.829E-01  &   5.351E-01\\
1.729  &   8.349E-02  &   1.489E-01  &  3.315E-01  &   5.639E-01\\
1.800  &   2.790E-02  &   1.868E-01  &  3.796E-01  &   5.943E-01\\
1.900  &   2.390E-02  &   1.824E-01  &  4.073E-01  &   6.136E-01\\
2.000  &   1.870E-02  &   1.823E-01  &  4.433E-01  &   6.443E-01\\
2.200  &   1.330E-02  &   1.851E-01  &  4.784E-01  &   6.768E-01\\
2.500  &   1.010E-02  &   1.939E-01  &  4.713E-01  &   6.753E-01\\
3.000  &   8.700E-03  &   2.202E-01  &  4.572E-01  &   6.862E-01\\
3.500  &   9.100E-03  &   2.398E-01  &  2.872E-01  &   5.361E-01\\
4.000  &   1.000E-02  &   2.497E-01  &  2.324E-01  &   4.922E-01\\
4.500  &   1.120E-02  &   2.638E-01  &  1.937E-01  &   4.688E-01\\
5.000  &   1.260E-02  &   2.797E-01  &  2.389E-01  &   5.312E-01\\
\noalign{\smallskip}
\hline
\end{tabular}
\end{table}